\documentclass[useAMS,usenatbib]{mn2e} 
\usepackage{aas_macros}
\usepackage{graphics}
\usepackage{times}
\usepackage{amssymb}

\def\healpix{{\sevensize HEALPIX}}

\def\camgrid{{\sevensize CAMGRID}}
\def\be{\begin{equation}}
\def\ee{\end{equation}}
\def\ba{\begin{eqnarray}}
\def\ea{\end{eqnarray}}
\def\nn{\nonumber}
\def\lgl{\langle}
\def\rgl{\rangle}

\def\vb{\bmath{b}}

\def\vn{\bmath{n}}
\def\vm{\bmath{m}}
\def\vd{\bmath{d}}

\def\lm{\ell m}

\def\vgamma{\bmath{\gamma}}
\def\vepsilon{\bmath{\epsilon}}
\def\mA{\mathbfss{A}}

\def\mP{\mathbfss{P}}
\def\mC{\mathbfss{C}}

\def\lsim{\mathrel{\rlap{\lower4pt\hbox{\hskip1pt$\sim$}}
    \raise1pt\hbox{$<$}}}                
\def\gsim{\mathrel{\rlap{\lower4pt\hbox{\hskip1pt$\sim$}}
    \raise1pt\hbox{$>$}}}
\def\gamia{\tilde{\gamma}^I}
\def\dgamia{\delta, \tilde{\gamma}^I}

\pagerange{\pageref{firstpage}--\pageref{lastpage}}
\pubyear{2010}

\begin{document}

\label{firstpage}

\title[Polarization as an indicator of intrinsic alignment in radio
  weak lensing]{Polarization as an indicator of intrinsic alignment in
  radio weak lensing}
  \author[M. L. Brown \& R. A. Battye]{Michael
  L. Brown$^{1,2}$\thanks{E-mail: mbrown@ast.cam.ac.uk} \& Richard
  A. Battye$^{3}$\thanks{E-mail: Richard.Battye@manchester.ac.uk}\\
  $^{1}$ Astrophysics Group, Cavendish Laboratory, University of
  Cambridge, J. J. Thomson Avenue, Cambridge CB3 OHE \\
  $^{2}$ Kavli Institute for Cosmology, University of Cambridge, Madingley
  Road, Cambridge CB3 OHA \\
  $^{3}$ Jodrell Bank Centre of Astrophysics, School of Physics and
  Astronomy, University of Manchester, Oxford Road, Manchester M13 9PL}
\date{\today} 

\maketitle

\begin{abstract}
  We propose a new technique for weak gravitational lensing in the
  radio band making use of polarization information. Since the
  orientation of a galaxy's polarized emission is both unaffected by
  lensing and is related to the galaxy's intrinsic orientation, it
  effectively provides information on the unlensed galaxy position
  angle. We derive a new weak lensing estimator which exploits this
  effect and makes full use of both the observed galaxy shapes and the
  estimates of the intrinsic position angles as provided by
  polarization. Our method has the potential to both reduce the
  effects of shot noise, and to reduce to negligible levels, in a
  model-independent way, all effects of intrinsic galaxy alignments.
  We test our technique on simulated weak lensing skies, including an
  intrinsic alignment contaminant consistent with recent observations,
  in three overlapping redshift bins. Adopting a standard weak lensing
  analysis and ignoring intrinsic alignments results in biases of 5-10\% in
  the recovered power spectra and cosmological parameters. Applying
  our new estimator to one tenth the number of galaxies used for the
  standard case, we recover both power spectra and the input cosmology
  with similar precision as compared to the standard case and with
  negligible residual bias, even in the presence of a substantial
  (astrophysical) scatter in the relationship between the observed
  orientation of the polarized emission and the intrinsic orientation.
  Assuming a reasonable polarization fraction for star-forming
  galaxies, and no cosmological conspiracy in the relationship between
  polarization direction and intrinsic morphology, our estimator
  should prove a valuable tool for weak lensing analyses of
  forthcoming radio surveys, in particular, deep wide field surveys
  with e-MERLIN, MeerKAT and ASKAP and ultimately, definitive radio
  lensing surveys with the SKA.
\end{abstract}


\section{Introduction}
\label{sec:intro}
The bending of light by mass inhomogeneities in the Universe results
in coherent distortions in the observed images of faint background
galaxies. In recent times, significant progress has been achieved in
measuring this weak gravitational lensing effect (see
e.g.\ \citealt{massey10} for a recent review). Weak lensing on
cosmological scales (or `cosmic shear') was first detected only ten
years ago \citep{wittman00, kaiser00, bacon00, vanwaerbeke00}, and has
since been measured with steadily improving precision by a large
number of groups (see e.g.\ \citealt{fu08} for the most recent
constraints from the CFHT Legacy Survey). Since the underlying physics
of weak lensing is clean, future cosmic shear measurements that
include distance information on the source galaxies, are considered to
be one of the most promising techniques for constraining the growth of
cosmological fluctuations over cosmic time. Cosmic shear therefore
offers great promise for constraining a variety of cosmological
parameters including the amplitude of the fluctuations, neutrino
masses and the properties of dark energy (e.g.\ \citealt{albrecht06, peacock06}).

In order to measure weak lensing, one must average over the observed
shapes of a sufficient number of background galaxies --- since one
does not know, a priori, the intrinsic shape of any one galaxy, the
lensing distortion cannot be recovered from a single galaxy image. If
one assumes that there is no correlation in the intrinsic shapes of
galaxies, then by averaging over a sufficient number of source
galaxies, one obtains an unbiased estimate of the distortion induced
by lensing.

There are two caveats to this technique. First, there is both
theoretical motivation (e.g.\ \citealt{crittenden01, catelan01,
  mackey02, jing02, hirata04}) and observational evidence
\citep{brown02, heymans04, mandelbaum06, mandelbaum09, hirata07,
  brainerd09} for believing that the intrinsic shapes of galaxies are
not completely random. Such intrinsic shape correlations will mimic a
lensing signal, introducing spurious power (the so-called
intrinsic-intrinsic or `II' term) into estimates of the
auto-correlations of the lensing distortion (or `shear')
field. Moreover, intrinsic correlations can introduce spurious
anti-correlations between the shear estimated from galaxies which are
widely separated in redshift (the so-called lensing-intrinsic
interference or `GI' term; \citealt{hirata04}). Of these two effects,
the GI term is potentially the most worrisome as it is not easily
removed from cosmic shear estimators.

Secondly, even if the intrinsic shapes of galaxies are truly random,
the dispersion in galaxy shapes introduces a shot noise term. Since
this dispersion is very much larger than the sought-after weak lensing
signal, one requires large numbers of background galaxies in order to
beat down this noise. For this reason, the vast majority of weak
lensing surveys to date have been conducted in the optical. In
particular, weak lensing in the radio band has lagged behind somewhat
due to the much smaller galaxy number density achieved in radio
surveys. However, the next generation of radio surveys such as the
imminent e-MERLIN\footnote{http://www.merlin.ac.uk/e-merlin} and LOFAR
\citep{morganti10} experiments, the SKA pathfinders, MeerKAT
\citep{booth09} and ASKAP \citep{johnston08}, and ultimately, the
SKA\footnote{http://www.skatelescope.org} itself, will be of
sufficient sensitivity to achieve a comparable source galaxy number
density to planned optical surveys.

The only significant measurement of cosmic shear in the radio band is
the work of \cite{chang04} who made a statistical detection in the VLA
FIRST survey. Recently, a further attempt to measure a lensing signal
was applied to data from the VLA and MERLIN \citep{patel10}. 
This latter work did not detect a significant lensing
signal precisely because of the small number density of galaxies
typically found in radio surveys. However, it was able to assess the
feasibility of doing so and also proposed that systematic effects
could be removed by observing the same patch of the sky in the radio
and optical wavebands. 

One aspect of radio surveys that is potentially useful for weak
lensing is the additional polarization information which often comes
for free in these surveys. In particular, all of the forthcoming
surveys mentioned above will include full polarization
information. Numerous authors have demonstrated that, in almost all
cases of astrophysical interest, the orientation of the polarized
emission from a source galaxy is unaffected by gravitational lensing
\citep{kronberg91, dyer92, faraoni93, surpi99, sereno05}. It follows
that if there exists a correlation (or anti-correlation) between the
intrinsic morphological orientation of a source and its polarized
emission, then the observed polarization provides information on the
unlensed source orientation. Such a relationship certainly exists for
quasars where the polarization is closely aligned with the radio jets
and this effect has already been exploited to measure gravitational
lensing using polarization observations of quasars \citep{kronberg91,
kronberg96, burns04}. The number counts of future deep radio surveys
will be dominated by quiescent star-forming galaxies rather than
AGN. Observations in the local Universe indicate that the orientation
of the polarized emission from these sources is also strongly
anti-correlated with the orientation of the galaxy (e.g.\ \citealt{stil09}).

In this paper, we construct a new weak lensing estimator which folds
in the extra information on the intrinsic orientation of galaxies
which can potentially be provided by radio polarization
observations. We demonstrate that, depending on the strength of the
correlation between the polarized emission and the galaxy orientation,
our proposed technique has the potential to reduce the impact of
shot noise and to mitigate intrinsic alignment (IA) effects. Focusing on the
potential impact for cosmic shear measurements, we test our technique
on simulated weak lensing skies which include an IA
contaminant. We demonstrate that IA biases can be
reduced to negligible levels, in a model-independent way, with
negligible loss in cosmological information as compared to a standard
weak lensing analysis which uses a factor ten more galaxies.

Note that our proposed technique is markedly different to other
techniques for mitigating the effects of IA which are generally based
on nulling the contaminating signals \citep{king02, heymans03,
  takada04, joachimi08, joachimi09a} or on fitting parametrized models
of the IA signals (e.g.\ \citealt{king03, king05, bridle07}). Some of the
nulling techniques are focused on removing the II term and some have
been constructed to explicitly remove the GI term, while the modeling
approaches attempt to model out all contaminating signals. In general,
the nulling techniques are lossy in the sense that they discard useful
cosmological information along with the IA contamination. The modeling
techniques are less lossy but are dependent on our highly uncertain
knowledge of the physics underlying IA effects. This model dependence
can be mitigated to some degree by including empirical constraints on
the IA signal, for example, by considering measurements of the cross
correlation between the galaxy number density and observed shear fields
\citep{zhang08, bernstein09, joachimi09b, kirk10}.

All of these existing techniques remove or model the IA contamination
at the two-point level (i.e.\ in the correlation functions or power
spectra). Since the technique proposed here works directly on the
shear field, it separates the IA and lensing signals at the map
level. It therefore does not discriminate between II and GI
contamination and can, in principle, be used to make maps (albeit
noisy ones) of the IA signal as well as to directly correct the
lensing shear field(s) for the effects of IA. In addition to cosmic
shear, in principle, our technique can also be used to correct for IA
contamination in other weak lensing applications, e.g.\ in lensing
reconstructions of dark matter distributions around clusters and
superclusters.

The paper is organized as follows. In Section~\ref{sec:pol_morph}, we
review the observational evidence for a correlation between the
orientation of the polarized emission and the intrinsic position angle
of galaxies. In Section~\ref{sec:estimator}, we derive our new
estimator and examine some of its properties. In
Section~\ref{sec:sims} we describe our simulations of weak lensing
skies including an IA contaminant. The analysis of the simulations is
presented in Section~\ref{sec:analysis}. We finish by discussing our
results and their implications for future radio lensing surveys in
Section~\ref{sec:discussion}.

\section{Polarization position angle as a proxy for the intrinsic position angle}
\label{sec:pol_morph} 

In this section we will discuss the use of the polarization position
angle as a proxy for the intrinsic structural position angle (PA) of a
radio source. At the low frequencies which we will consider (typically
1-10 GHz), the dominant source of linear polarization is expected to
be synchrotron radiation due to electrons moving in 
magnetic fields. A measurement of the polarization position angle (PPA) gives information
about the direction of the magnetic fields and, as we will discuss below, it is reasonable to
suppose that this is related to the overall structure of the
galaxy. The PPA measurement  can be confused by Faraday rotation which would need to
be extracted in any observations by {\it Rotation Measure Synthesis}
using observations covering a wide range of frequencies (for example,
as described in \citealt{beck04}).

The linear polarization of a source is usually described in terms of
the Stokes parameters $Q$ and $U$ and the PPA, $0^{\circ}\le\alpha\le
180^{\circ}$, is conventionally defined as
$$
\alpha={1\over 2}{\rm tan}^{-1}\left({U\over Q}\right)\,.
$$
If the noise on the measurement of $Q$ is $\sigma_{Q}$ and that on
$U$ is $\sigma_{U}$, then if $\sigma_{Q}=\sigma_{U}=\sigma$ one can
show that the r.m.s. error on the measurement of $\alpha$ due to the
instrument noise will be $\langle\Delta\alpha^2_{\rm
  N}\rangle^{1/2}=\sigma/(2P)$ where $P=\sqrt{Q^2+U^2}$ is the polarized
intensity of the source. For a SNR (signal-to-noise ratio) of 3 this
corresponds to $\approx 10^{\circ}$ and for a SNR of 5 it is $\approx
6^{\circ}$. We note that for low values of the SNR the probability
distribution function will be non-Gaussian. For higher SNRs one would
expect any intrinsic dispersion to dominate. We will quantify this by
the variance $\langle\Delta\alpha^2_{\rm int}\rangle$ which we will
assume is independent of the instrument noise and hence adds in
quadrature.

The parameters which it is important to quantify in order to assess
the viability of our method relative to standard techniques are the
mean error in the position angle estimator,
$\langle\Delta\alpha\rangle$, its total r.m.s. scatter, $\alpha_{\rm
  rms}=\langle\Delta\alpha^2\rangle^{1/2} = \sqrt{(\sigma^2/4P^2) + 
  \langle\Delta\alpha^2_{\rm int}\rangle}$, and the distribution of 
fractional polarization $\Pi=P/I$ for total intensity $I$, quantified
by the median fractional polarization $\Pi_{\rm med}$ and possibly
some scatter. We will assume that $\langle\Delta\alpha\rangle=0$ which
will ensure that our estimator is unbiased in the absence of intrinsic
alignments; although there are many possible sources of astrophysical
scatter in such measurements, it seems implausible that any of these
will prefer any particular direction.

The emission from radio galaxies which is relevant to us here can be
divided into two broad classes --- star-forming galaxies and Active
Galactic Nuclei (AGN) --- and the nature of polarization will be very
different in both. The expectation is that in each case the
polarization position angle will be perpendicular to the intrinsic
axis of the galaxy, albeit for different reasons. The category of AGN
is often sub-divided into radio-loud, jet powered (FRI/FRII) and
radio-quiet quasars. Both are often connected with elliptical
morphologies. Star-forming galaxies are often categorized as either
quiescent ``normal'' galaxies, such as our own, and starbursts which
are undergoing an epoch of intensive star-formation. The star-forming
galaxies typically have spiral morphologies. AGN are thought to
dominate the source counts for $I>1\,{\rm mJy}$ and star-forming
galaxies for $I<10\,\mu{\rm Jy}$ with the transition expected to
happen somewhere in between. We are interested in situations where the
source density is high ($\sim 20$ arcmin$^{-2}$) in order for it to be
possible to measure cosmic shear and hence we are interested in the
sources with $I\approx1\,\mu{\rm Jy}$. Therefore we are most
interested in the properties of star-forming galaxies where the
detected radio flux is dominated by emission from massive star
formation. Unfortunately, very little is known about the polarization
properties of such objects since they are a very small fraction of the
sources detected in presently available surveys. In what follows we
will attempt to piece together a picture of what is known about this
population from the information available in the literature.

\cite{wilman08} have constructed a simulated catalogue of radio
sources as part of the Square Kilometre Array Design Studies (SKADS)
program.\footnote{available at http://s-cubed.physics.ox.ac.uk} This
is a semi-empirical simulation of the extragalactic radio continuum
sky over a $20 \times 20$ deg$^2$ area which includes FRI, FRII,
radio-quiet quasars, ``normal'' and star-burst galaxies. At levels of
$I\approx 1\,\mu{\rm Jy}$ the radio source population is dominated by
star-forming (``normal'' or star-burst) galaxies. For the central
square degree of the simulation we find that there are $\approx 80\%$
star-forming galaxies ($\approx 72.5\%$ ``normal'' and $\approx 7.5\%$
star-burst galaxies) and $\approx 20\%$ AGN. This is $\approx 50\%$ of
each at $I\approx 50\,\mu{\rm Jy}$ and the proportion of star-forming
galaxies increases for lower flux densities. For a detection threshold
of $1\,\mu{\rm Jy}$, the median redshift of the star-forming galaxies
is $\approx 2$. The AGN types will act as a contaminating noise
background when attempting to use the PPA as a proxy for the
structural PA since the radio emission is largely dominated by the
jets in such sources and the alignment properties between the
polarization position angle and the structural axis will typically be
different, although still aligned \citep{clarke80}. It should be
possible to remove this contaminating population using a combination
of indicators. These will include morphological information on the
source, the use of other radio observations, for example, spectral
index information or 21cm line measurements, or information garnered
from observations at other wavebands.

\begin{figure*}
  \centering
  \resizebox{0.75\textwidth}{!}{  
    \rotatebox{0}{\includegraphics{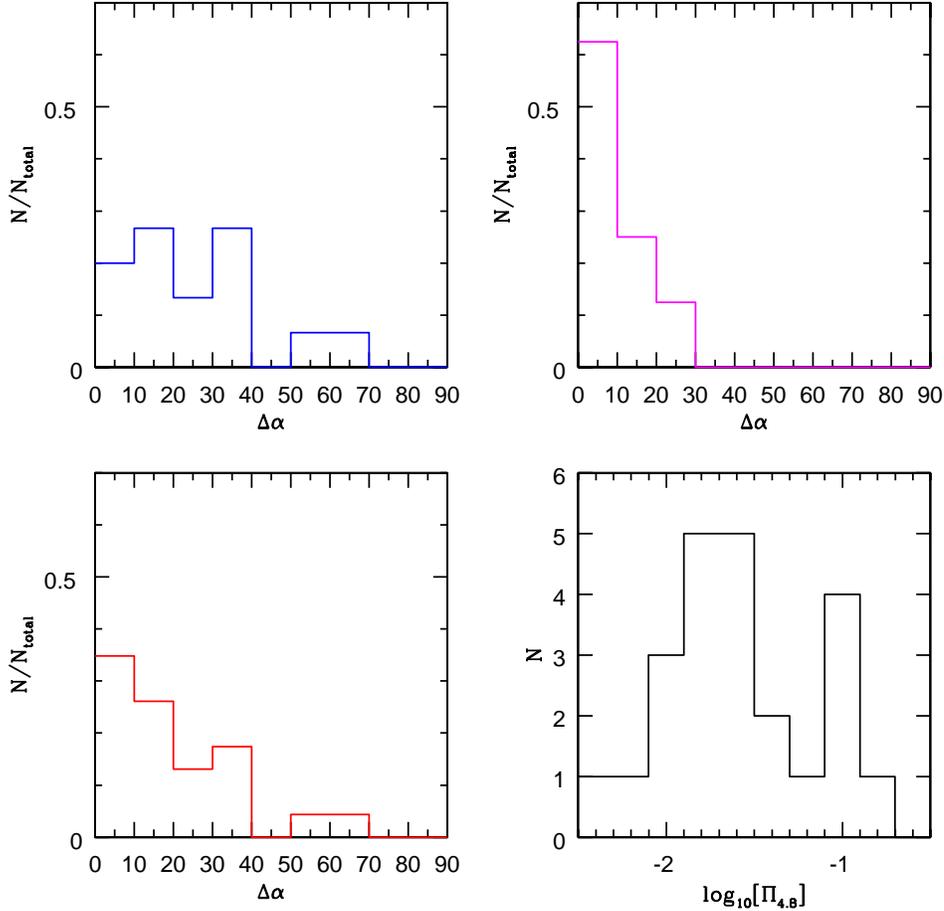}}}
  \caption{A summary of the results from {\protect \cite{stil09}}.
    \emph{Bottom left:} Histogram of the number of objects divided by
    the number in the sample as a function of $\Delta\alpha$ for the
    nearby spiral and Virgo samples of galaxies. \emph{Bottom right:}
    Histogram of the number of objects as a function of
    $\log\Pi$. \emph{Top left:} The equivalent of the bottom left
    figure but for $\Pi_{4.8}<0.03$. \emph{Top right:} The equivalent
    of bottom left figure but for $\Pi_{4.8}>0.03$.}
  \label{fig1}
\end{figure*}

\cite{stil09} have performed a detailed study of star-forming galaxies
in the local Universe. They extracted information from radio
observations of a number of local spiral galaxies at 4.8 and 8.4 GHz
classifying them into three sub-samples: standard nearby spirals,
Virgo (in the Virgo Cluster) and barred spirals. They computed the
PPAs and compared them to corresponding structural PAs deduced in the
optical. They found a strong anti-correlation between the PPA and
structural PA for the standard nearby spiral and Virgo samples which
is illustrated in Fig.~\ref{fig1} (bottom left) where we present a
histogram of the number of objects divided by the total number in the
sample as a function of the acute angle ($\Delta\alpha$) between the
optical PA and the direction perpendicular to the radio PPA at
$4.8\,{\rm GHz}$. This has an r.m.s. of
$\langle\Delta\alpha^2\rangle^{1/2}\approx 26.7^{\circ}$. The
contribution due to this from random errors is expected to be
negligible and therefore we consider this as an upper limit on
$\langle\Delta\alpha_{\rm int}^2\rangle^{1/2}$ provided we assume that
the population at high redshifts is dominated by similar galaxies. The
situation is much less clear for the barred spirals.

For our purposes we actually require the difference between the
perpendicular to the PPA in the radio and the structural PA deduced
from the total intensity distribution detected in the
radio. \cite{battye09} have investigated the differences between
structural PAs measured in the optical from the SDSS \citep{york00}
and in the radio from FIRST \citep{becker95}. They found that, for
star-forming galaxies identified photometrically or spectroscopically,
the data were compatible with an r.m.s. difference which is $\approx
15^{\circ}$. Some of the dispersion observed by \cite{stil09} could be
due to them using the optical PA and this could be responsible for the
less clear correlation in the barred galaxies whose estimated optical
PAs could be very different to their radio PA. Under the assumption
that the PPA observed in the radio is more strongly anti-correlated with
the structural PA observed in the radio than it is with the structural
PA observed in the optical, and assuming that the two effects are
independent, we estimate $\langle\Delta\alpha^2_{\rm
int}\rangle^{1/2}<22.5^{\circ}$.

The fractional polarization of these galaxies ($\Pi_{\rm med}=0.024$)
is typically higher than that found for AGN type galaxies. We present
a histogram of the fractional polarization of the standard nearby
spiral and Virgo samples in Fig.~\ref{fig1} (bottom right). There
appears to be a peak around $\Pi\approx 0.1$ which might suggest a
population of sources with high fractional polarization although this
is not statistically significant. We have produced histograms of
$\Delta\alpha$ for $\ \Pi_{4.8}<0.03$ and $\Pi_{4.8}>0.03$ which are
presented in Fig.~\ref{fig1} (top left and top right,
respectively). For low fractional polarization there is still a trend
for $\Delta\alpha$ to be biased toward zero but with a higher
dispersion $\langle\Delta\alpha^2\rangle^{1/2}\approx 31.7^{\circ}$,
whereas for higher fractional polarization the histogram has a
dispersion of only $\langle\Delta\alpha^2\rangle^{1/2}\approx
13.2^{\circ}$. This latter value would be compatible with
$\langle\Delta\alpha^2_{\rm int}\rangle\approx 0^{\circ}$ if the
structural PA is measured from the total intensity distribution
observed in the radio. \cite{stil09} also report an
anti-correlation between fractional polarization and the luminosity
measured at 4.8 GHz. They also comment that high fractional
polarization is often connected with high inclination angles of
the associated disk. For example, the 4 sources with the highest
reported fractional polarization at 4.8 GHz have inclination angles in
the range $75-90^{\circ}$.

The anti-correlation between the structural PAs deduced in the radio
and optical and the PPA is to be expected in star-forming galaxies
since radio and optical emission are both dominated by the stars in
the disk of the galaxy, albeit from somewhat different star
populations. The massive star-formation and the magnetic field
responsible for the radio emission should be aligned with the galaxy's
disk and the higher the level of alignment, the higher the fractional
polarization is expected to be. This, and the observational evidence
discussed in the previous paragraph, leads to the interesting
conclusion that the level of intrinsic scatter in $\Delta\alpha$ will
be anti-correlated with $\Pi$. This means that galaxies with higher
fractional polarization, in which the polarization is easier to detect
relative to lower fractional polarization counterparts, have lower
values of $\langle\Delta\alpha_{\rm int}^2\rangle$. Effectively, a
sub-sample of galaxies selected to have high fractional polarization
will have a low dispersion in the position angle proxy relative to the
whole sample. Note however that such a sub-sample might also suffer
from an enhanced dispersion in intrinsic ellipticity since highly
inclined systems would be preferentially selected.

There is some evidence that the fractional polarization is increasing
at low flux densities at 1.4 GHz \citep{taylor07, subrahmanyan10,
  grant10}. This increase appears to start around
$I\approx 10\,{\rm mJy}$ and continues at least down below $I\approx
1\,{\rm mJy}$. There is general agreement that some increase takes place
but disagreement about how much, although we note that the results are
not incompatible since they probe different regimes of flux density
and that the discrepancies are not that significant in the regions
where they overlap. \cite{taylor07} and \cite{grant10} find
$\Pi_{\rm med}\approx 0.05$ for $I\approx 10\,{\rm mJy}$, compared to
$\Pi_{\rm med}\approx 0.015$ for sources with high flux density,
whereas \cite{subrahmanyan10} suggest $\Pi_{\rm med}\approx 0.15$
below $I\approx 1\,{\rm mJy}$. Since it is thought that the star-forming
galaxies start to become a significant fraction of the sources around
these flux densities, it is tempting to believe that this rise,
however big it might be, is at least partially due to the population
of star-forming galaxies although it could equally likely be due to a
hitherto unidentified population of AGN with high fractional
polarization.

In order to make an accurate measurement of the ellipticity from the
total intensity will require a high SNR detection since it requires
the measurement of 3 parameters --- the semi-major and minor axes and
the position angle, or equivalently the quadrupole
moments. \cite{blake07} have estimated that this will require $\approx
10\sigma$ detections. Accurate measurement of the polarization position
angle, such that the random error is comparable to the scatter
expected from the work of \cite{stil09}, probably only requires a
$3-5\sigma$ detection. If, as we have attempted to argue, the median
level of fractional polarization of the star-forming galaxies is
$\Pi_{\rm med}\sim 0.1$ then we will get useful polarization
information from $\sim 20-30\%$ of the galaxies for which an accurate
measurement of the ellipticity is possible.

Clearly there is still some uncertainty in the predictions discussed
above. Probably, the most important is that the population which will
be probed by observations relevant to weak lensing will be at
substantially higher redshifts than that discussed in
\cite{stil09}. At these high redshifts even ``normal galaxies'' will
be undergoing substantial star-formation and hence the radio and optical
emission, and the orientation of the polarized emission may not be as
well aligned as in the local Universe. This will be investigated as
part of future observational programs leading up to the SKA.

The evidence we have discussed above presents a prima-facie case that
$\langle\Delta\alpha^2_{\rm int}\rangle^{1/2}$ is somewhere between
$0^{\circ}$ and about $20^{\circ}$ and that $\Pi_{\rm med}$ could
be as large as 0.1. In addition we have argued on physical grounds
that $\langle\Delta\alpha^2_{\rm int}\rangle^{1/2}$ is anti-correlated
with $\Pi$ implying that we will preferentially select objects which
are more aligned and that they can possibly be weighted by their
fractional polarization. These assertions are, of course, uncertain
and the amount of information available at present is clearly
insufficient to make any strong statement. In what follows we will
attempt to show how our methods depend on $\alpha_{\rm rms}$ and the
number of galaxies for which we can make an accurate polarization
measurement. However, for the simulations of Section~\ref{sec:sims},
we are forced to pick specific values and, when necessary, we will
assume that $10\%$ of the galaxies detected by the SKA will be
sufficiently well detected for there to be a total r.m.s. uncertainty
in the position angle of $\alpha_{\rm rms}=5^{\circ}$. This seems to be a
reasonable balance within the range of possibilities which are
discussed above.

\section{Shear estimation with an intrinsic position angle estimate}
\label{sec:estimator}
The relationship described in the previous section between the
intrinsic position angle of a source galaxy and the orientation of its
polarized emission effectively provides us with an estimate, however
noisy, of the intrinsic position angle of the galaxy. In order to fold
this information into a weak lensing analysis, a new estimator is
required which takes full advantage of the available information. In
this section, we derive the appropriate estimator and present some of
its properties. 

\subsection{Derivation}
\label{sec:estimator_derivation}
The effect of weak lensing on a background galaxy's ellipticity is 
\be
 \vepsilon^{\rm obs} = \vepsilon^{\rm int} + \vgamma,
\label{eqn:wl_basic}
\ee
where $\vepsilon = \epsilon_1 + i \epsilon_2$
is the (complex) ellipticity and $\vgamma =
\gamma_1 + i \gamma_2$ is the shear. The superscripts
`obs' and `int' denote the observed and intrinsic ellipticity
respectively. Here we've used a definition of the ellipticity of: 
\be
\vepsilon = \frac{Q_{11} + Q_{22} + 2iQ_{12}}{Q_{11}
+ Q_{22} + 2(Q_{11}Q_{22} - Q^2_{12})^{1/2}},
\label{eqn:ellip_defn}
\ee
where the $Q$'s are the (weighted) quadrupole moments of the galaxy
image. Note that, depending on the definition of ellipticity used,
equation~(\ref{eqn:wl_basic}) sometimes appears with a factor 2 multiplying
the shear (see e.g.\ \citealt{bartelmann01} for a discussion). We can
also write the ellipticity (or shear) in polar form, e.g.,
\be
\vepsilon = | \epsilon | \exp(i \, 2\alpha),
\label{eqn:ellip_polar}
\ee where $ | \epsilon |^2 = \epsilon^2_1 + \epsilon^2_2$ is the
amplitude of the ellipticity and $\alpha$ is the orientation. We also
have $\epsilon_1 = | \epsilon | \cos(2 \alpha)$ and $\epsilon_2 = |
\epsilon | \sin(2 \alpha)$.

The standard estimator for the average shear field in a pixel on the
sky is simply to take the average of the observed galaxy ellipticities:
\be
\hat{\vgamma} = \frac{1}{N} \sum^N_{i = 1} \vepsilon_i^{\rm obs},
\label{eqn:standard_est}
\ee
where the sum is over all galaxies falling within this pixel. From
equation~(\ref{eqn:wl_basic}), the expectation value of this estimator is 
\be
\lgl\hat{\vgamma}\rgl = \vgamma + \lgl \vepsilon^{\rm int} \rgl. 
\label{eqn:standard_est_expval}
\ee
In the absence of an IA signal, $\lgl \epsilon^{\rm int} \rgl = 0$
and this estimator is an unbiased estimator for the average shear in a
pixel. The dispersion is
\ba
\sigma^2_{\hat{\vgamma}} & = & ( \lgl \hat{\vgamma} \hat{\vgamma}^*
\rgl - \lgl \hat{\vgamma} \rgl \lgl \hat{\vgamma} \rgl^* ) / N \nn \\
& = & \lgl \vepsilon^{\rm int} {\vepsilon^{\rm int}}^* \rgl / N,
\label{eqn:standard_est_error}
\ea
where the asterix denotes complex conjugation. In standard weak
lensing analyses, this term is the irreducible shot noise due to the
dispersion in the (intrinsic) shapes of galaxies.

Writing equation~(\ref{eqn:wl_basic}) in component form and in terms of the
intrinsic ellipticity amplitude and position angle, we have:
\ba
\epsilon^{\rm obs}_1 = |\epsilon^{\rm int}| \cos(2\alpha^{\rm int}) + \gamma_1 \nn \\
\epsilon^{\rm obs}_2 = |\epsilon^{\rm int}| \sin(2\alpha^{\rm int}) + \gamma_2.
\label{eqn:wl_basic_comp}
\ea
The new information coming from the orientation of the polarized
emission effectively gives us a noisy estimate of the intrinsic
position angle, $\alpha^{\rm int}$.  Even with this extra information, for
a single galaxy, there are two equations and three unknowns
($\gamma_1, \gamma_2$ \& $|\epsilon^{\rm int}|$) and so there is no unique
solution. However, if we pixelise the sky and assume that the lensing
shear field is approximately constant within each pixel, then all we
need is at least two galaxies in each pixel to be able to solve the
system. In general, if we have $N$ galaxies in a pixel, for each of
which we have estimates of the ellipticity components
($\epsilon^{\rm obs}_{1,2}$) and intrinsic position angle
($\hat{\alpha}^{\rm int}$), then we have for each galaxy:
\ba
\epsilon^{\rm obs}_{1,i} = |\epsilon_i^{\rm int}| \cos(2\hat{\alpha}_i^{\rm int}) + \gamma_1 \nn \\
\epsilon^{\rm obs}_{2,i} = |\epsilon_i^{\rm int}| \sin(2\hat{\alpha}_i^{\rm int}) + \gamma_2.
\label{eqn:wl_basic_comp_ngal}
\ea
Now, we have $2N$ equations and $N+2$ unknowns ($\gamma_1, \gamma_2$
and $|\epsilon_i^{\rm int}|$ for $i = 1,...,N$). The system is therefore
well-defined and soluble --- in principle, exactly. Taking the ratio
of the two equations in (\ref{eqn:wl_basic_comp_ngal}) and re-arranging, we have 
\ba 
\epsilon^{\rm obs}_{1,i} \sin(2\hat{\alpha}_i^{\rm int}) \!\!\!\!&-&\!\!\!\!
\epsilon^{\rm obs}_{2,i} \cos(2\hat{\alpha}_i^{\rm int}) \nn \\
\!\!\!\!&=&\!\!\!\! \gamma_1 \sin(2\hat{\alpha}_i^{\rm int}) - \gamma_2
\cos(2\hat{\alpha}_i^{\rm int}) 
\label{eqn:wl_ratio}
\ea
Defining the pseudo-vectors, 
\be
\hat{\vn}_i = \left( \begin{array}{c} \sin 2\hat{\alpha}_i^{\rm int} \\ -\cos
  2\hat{\alpha}_i^{\rm int} \end{array} \right) \!;
\, {\vepsilon}_i^{\rm obs} = \left( \begin{array}{c} \epsilon_{1,i}^{\rm obs}
  \\ \epsilon_{2,i}^{\rm obs} \end{array} \right) \!;
\, {\vgamma} = \left( \begin{array}{c} \gamma_1
  \\ \gamma_2 \end{array} \right) \!,
\label{eqn:pseudo_vecs}
\ee
equation~(\ref{eqn:wl_ratio}) can be written in the more compact form,
\be
\hat{\vn}_i \cdot \vepsilon_i^{\rm obs} = \hat{\vn}_i \cdot \vgamma.
\label{eqn:wl_ratio_compact}
\ee
Note that the vector, $\hat{\vn}_i$, is simply the direction which is
at $45^\circ$ to our estimate of the intrinsic position angle for each
galaxy. 

We wish to select our estimate of the shear such that the constraint
of equation~(\ref{eqn:wl_ratio_compact}) is enforced for each galaxy
in our pixel. To achieve this, we define a $\chi^2$ for the shear in
each pixel as 
\be 
\chi^2 = \sum_i w_i \left[ \hat{\vn}_i \cdot ( \vepsilon_i^{\rm obs} -
  \vgamma) \right]^2,
\label{eqn:wl_chi2}
\ee
where the sum is over all galaxies in a pixel and $w_i$ is an
arbitrary weight assigned to each galaxy (which should be normalized
to unity). Minimizing equation~(\ref{eqn:wl_chi2}) with respect to $\vgamma$
gives us our new estimator for the average shear in a pixel which we
can write as
\be
\hat{\vgamma} = \mA^{-1} \vb,
\label{eqn:new_est}
\ee
where the matrix, $\mA$, and the vector, $\vb$, are given by
\ba
\mA &=& \sum_i w_i \hat{\vn}_i \hat{\vn}^T_i, 
\label{eqn:Amatrix_def} \\
\vb &=& \sum_i w_i (\vepsilon_i^{\rm obs} \cdot \hat{\vn}_i) \hat{\vn}_i.
\label{eqn:bvec_def}
\ea
The matrix, $\mA$ is simply the weighted sum of the projection
matrices, $\mP_i = \hat{\vn}_i \hat{\vn}_i^T$. In minimizing the
$\chi^2$ of equation~(\ref{eqn:wl_chi2}), our estimator effectively
selects the unique shear, $\hat{\vgamma}$ such that the average
difference between it and the observed ellipticities, $\vepsilon_i$,
projected in a direction which is at $45^\circ$ to our estimates of
the intrinsic position angles, is minimized.

Alternatively, if for each galaxy we rotate the coordinate
system such that it is aligned with our estimate of the intrinsic
position angle, then the components of the observed ellipticity in
this rotated (local) coordinate frame are simply 
\ba
\epsilon^{\rm local}_{1,i} &=& \hat{\vn}^{\|}_i \cdot \vepsilon_i^{\rm obs}, \nn \\
\epsilon^{\rm local}_{2,i} &=& \hat{\vn}_i \cdot \vepsilon_i^{\rm obs},
\label{eqn:e1_e2_local}
\ea
where $\hat{\vn}^{\|}_i = \left( \cos2\hat{\alpha}_i^{\rm int},
\sin2\hat{\alpha}_i^{\rm int} \right)$ is the direction parallel to the
estimate of the intrinsic position angle. The estimator of
equation~(\ref{eqn:new_est}) thus retains the $\epsilon^{\rm local}_2$
component and discards the $\epsilon^{\rm local}_1$ component of each
galaxy. 

Note that one has complete freedom in choosing the weights, $w_i$. For
example, these could be chosen in order to downweight galaxies with
low signal-to-noise polarization measurements, low fractional
polarization, or to give more weight to highly regular objects whose
polarization properties are extremely well understood.

\subsection{Properties of the estimator}
If our estimates of the intrinsic position angles
($\hat{\alpha}^{\rm int}_i$) were perfect, then it is easy to show that
the estimator of equation~(\ref{eqn:new_est}) is shot noise free
and would also eliminate any and all IA effects. In
this case, all one would require for a sample-variance limited lensing
survey insensitive to IA contamination would be two galaxies with
significantly different intrinsic position angles in each sky pixel.

However, as described in Section~\ref{sec:pol_morph}, there will be
an irreducible astrophysical scatter in the relationship between the
true intrinsic position angle and that inferred from the observed
orientation of the polarized emission. This scatter will re-introduce
shot noise into our estimator, the level of which is determined by the
size of the scatter in $\hat{\alpha}^{\rm int}_i$ and the number density of
galaxies. In the presence of an IA signal, this scatter will also
result in a small residual bias in the estimator. 

\begin{figure*}
  \centering
  \resizebox{0.90\textwidth}{!}{  
    \rotatebox{0}{\includegraphics{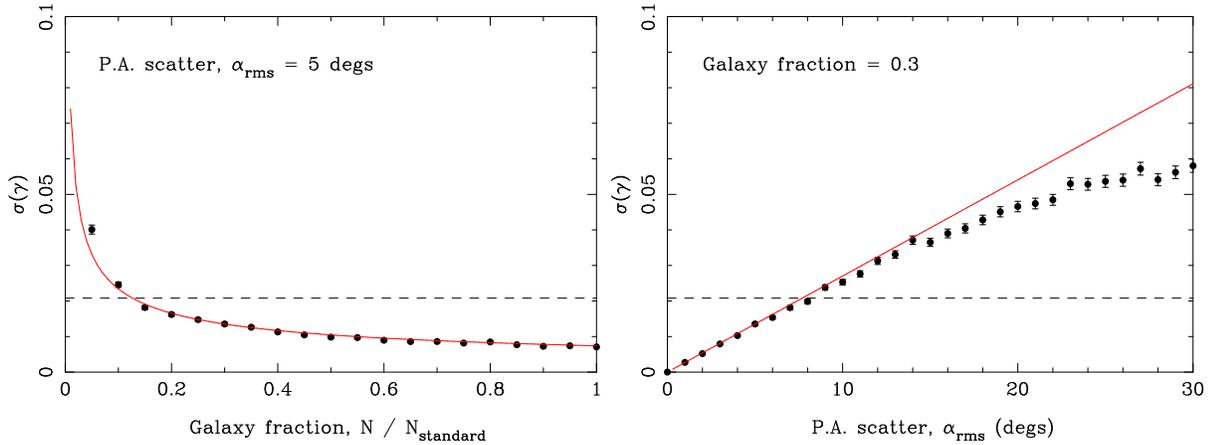}}}
  \caption{\emph{Left panel:} Dispersion of the estimator (equation
    \ref{eqn:new_est}) as a function of galaxy number density for a
    scatter in the intrinsic position angle estimates of 5
    degs. \emph{Right panel:} The dispersion as a function of the
    scatter in the intrinsic position angle estimates for a galaxy
    number density of 30\% of that used in the standard case. In both
    cases, the curves shows the analytic approximation of
    equation~(\ref{eqn:new_est_error}), the points are the
    result from a numerical calculation and the dashed horizontal
    lines show the dispersion of the standard estimator (for $N =
    N_{\rm standard}$) for comparison.}
  \label{fig2}
\end{figure*}

We show in Appendix~\ref{app:estimator_properties} that, in the
absence of IA signals, and for small errors in the
estimates of the intrinsic position angles, $\alpha_{\rm rms} \ll 1$,
the standard error is well approximated by 
\be
\sigma_{\hat{\vgamma}} \approx 4 \frac{\alpha_{\rm rms} 
\, \vepsilon_{\rm rms}}{\sqrt{N}},
\label{eqn:new_est_error}
\ee 
where $\vepsilon_{\rm rms} = \lgl \vepsilon^{\rm int} 
{\vepsilon^{\rm int}}^* \rgl^{1/2}$ is the dispersion in intrinsic
ellipticities. Fig.~\ref{fig2} shows a comparison between this
approximation and the exact numerical result. Clearly, for larger
scatters ($\gsim 10$ degs) in the intrinsic position angle estimates,
the approximation breaks down and overestimates the true error in the
estimator.
In the absence of IA signals, the main impact our technique
would have is in reducing the shot noise in lensing
reconstruction. For example, for $\alpha_{\rm rms} = 0.08$, which
corresponds to a $4.6^\circ$ dispersion in the intrinsic position
angle estimates, one would require a factor 10 less galaxies than in
the standard case to achieve the same level of shot noise.

\begin{figure*}
  \centering
  \resizebox{0.90\textwidth}{!}{  
    \rotatebox{0}{\includegraphics{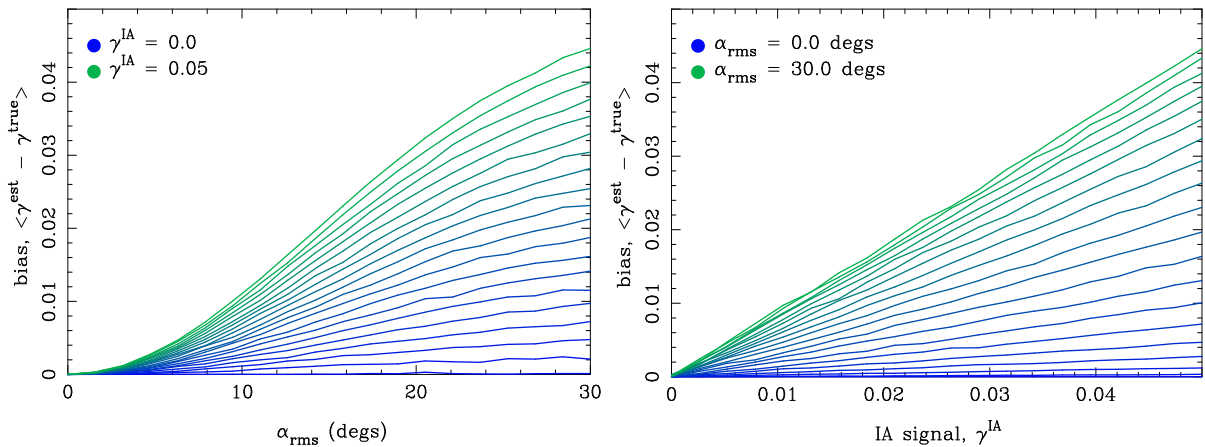}}}
  \caption{Residual bias in the estimator of
  equation~(\ref{eqn:new_est}) in the presence of an IA signal and
  with a non-zero error in the estimates of the intrinsic position
  angles. The left hand panel shows the bias as a function of the
  intrinsic position angle scatter for monotonically increasing IA
  signals. The right hand panel shows how the bias scales with the IA
  signal for monotonically increasing values of $\alpha_{\rm rms}$.}
  \label{fig3}
\end{figure*}

In the presence of IA signals, errors in the intrinsic position angle
estimates will mean that the contamination is not completely removed,
that is, the estimator is biased. However, we find that this bias is
very much smaller than for the standard estimator. A general
expression for the residual bias is hard to come by since this bias
depends on the intrinsic position angles of all the galaxies in a
pixel. One therefore needs to rely on numerical simulations but we
note that the residual bias will depend on the scatter in the
intrinsic position angle estimates, $\alpha_{\rm rms}$ and on the IA
signal itself. We note further that the residual bias is independent
of both the number density of galaxies and the lensing shear
signal. Figure~\ref{fig3} shows the bias in the recovered shear as
measured from numerical simulations as a function of both $\alpha_{\rm
rms}$ and the IA signal. For an intrinsic position angle scatter of
$\alpha_{\rm rms} = 5^\circ$, the bias due to IA effects is reduced by
over an order of magnitude compared to the bias found in the standard
estimator.

Note finally that we can construct an estimator for the IA
signal trivially using the standard and new estimators: 
\be
\hat{\vgamma}^{\rm IA} = \hat{\vgamma}^{\rm standard} -
\hat{\vgamma}^{\rm new}.
\label{eqn:ia_est}
\ee
In principle, such estimates could be used as a starting point to
iteratively correct for the residual bias in our shear estimator
although our simulations indicate that such a correction is unlikely
to be necessary provided that the position angle estimates from
polarization are good to $\sim 5^{\circ}$ (see also
Section~\ref{sec:bias_correction}). 

\section{Simulations}
\label{sec:sims}
In this section we will test our new estimator on simulated weak
lensing skies including a model of the IA signal. Our purpose here is
to demonstrate that, given an estimate of the intrinsic position angle
via the polarization direction, our technique can be a potentially
powerful way to minimize the impact of IA contamination in cosmic
shear surveys. We have therefore not performed detailed simulations of
a radio weak lensing survey. In particular, we do not consider
systematics associated with either the instrument or the
atmosphere. For a comprehensive description of the possible
systematics in radio surveys and their implications for weak lensing,
see \cite{chang04}. We simply note that some systematics are worse in
the radio (e.g.\ ionospheric distortions), whereas in other cases the
reverse is true (e.g.\ complicated optical point spread functions
versus precisely determined radio interferometer beam shapes).

We further restrict ourselves to pure Gaussian fields and ignore the
non-Gaussianity of the lensing shear field on small scales. Since we
consider the reconstruction of the lensing signal on medium to large
scales only (multipoles, $\ell \lsim 2000$), this should not
significantly affect our conclusions. 

\subsection{Background cosmology and survey parameters}
\label{sec:cosmo_model}
All of our simulations are generated within a background cosmology
based on the $\Lambda$CDM cosmological model with parameters,
$\Omega_m = 0.262$, $\sigma_8 = 0.798$, $H_0 = 71.4$ km s$^{-1}$
Mpc$^{-1}$ and $\Omega_b = 0.0443$. We additionally assume a flat
Universe, $\Omega_\Lambda = 1 - \Omega_m$.  

In what follows, we will be simulating weak lensing and IA fields at
multiple redshifts, including all possible cross-correlations between
the fields at different redshifts. To limit the complexity of the
system, we have therefore restricted our simulations to consider only
three broad redshift bins: $0.00 < z_1 < 1.40$, $1.40 < z_2 < 2.60$
and $z_3 > 2.60$. The bin limits were chosen such that each bin
contained approximately the same number density of sources. It is
likely that this choice would be sub-optimal for constraining
cosmological parameters in the analysis of a real radio lensing data
set but we have not investigated the optimal choice in this work.

As described in \cite{hirata04}, the various auto- and cross-power
spectra of the weak lensing and IA fields are given by
\ba
C^{GG}_{\ell(ij)} &=& \int_0^{\infty} \frac{W_i(\chi)
  W_j(\chi)}{\chi} P_{\delta}(k, \chi) \, d\chi \label{eqn:cl_gg} \\
C^{II}_{\ell(ij)} &=& \int_0^{\infty} \frac{f_i(\chi)
  f_j(\chi)}{\chi} P_{\gamia}(k, \chi) \, d\chi \label{eqn:cl_ii} \\
C^{GI}_{\ell(ij)} &=& \int_0^{\infty} \frac{W_i(\chi)
  f_j(\chi)}{\chi} P_{\dgamia}(k, \chi) \, d\chi \label{eqn:cl_gi},
\ea
where $k = \ell / \chi$. Here, $f_i$ is the normalized comoving
distance distribution of galaxies in bin $i$ and $W_i(\chi)$ is the
lensing selection function for the source galaxy distribution of bin
$i$. For a lens at a comoving distance, $\chi_d$ and with redshift,
$z_d$, this latter function can be written as
\be
W_i(\chi_d) = \frac{3}{2} \Omega_m \frac{H_0^2}{c^2} (1 + z_d)
\int_0^\infty f_i(\chi_s) \frac{(\chi_s - \chi_d)}{\chi_s} \, d\chi_s,
\label{eqn:wl_selection_fn}
\ee 
where the integration is over the distance to the source galaxies,
$\chi_s$. 

In equation~(\ref{eqn:cl_gg}), $P_{\delta}(k, \chi)$ is the normal 3D
matter power spectrum. To calculate this, we use the transfer function
fitting formulae of \cite{eisenstein99} and we use the {\sevensize
HALOFIT} code \citep{smith03} to calculate the nonlinear power
spectrum. The two other power spectra, $P_{\gamia}(k, \chi)$ and
$P_{\dgamia}(k, \chi)$ are the (projected) IA power spectrum and the
cross-power spectrum of the matter and IA fields respectively. These
functions are not well understood, either theoretically or
observationally, which motivates a model-independent technique for
removing the IA contamination. The analysis we present is such a
model-independent technique but we will still need to choose an IA
model to generate our simulated data sets. This is addressed in the
following section.

\begin{figure}
  \centering
  \resizebox{0.45\textwidth}{!}{  
    \rotatebox{-90}{\includegraphics{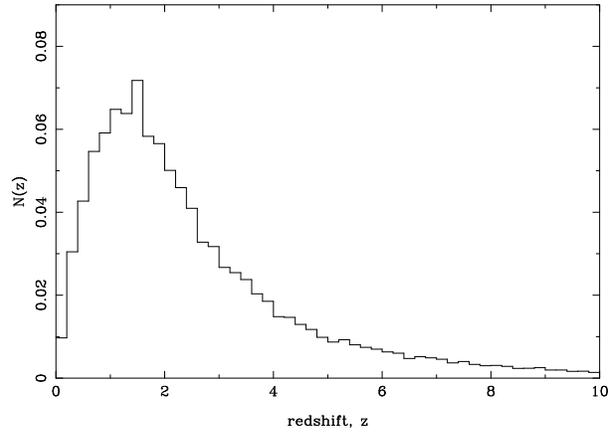}}}
  \caption{Normalized redshift distribution of star-forming galaxies in the
    SKADS simulations of {\protect \cite{wilman08}} down to a flux
    threshold of S$_{\rm 1.4 GHz} = 0.5\,\mu$Jy. We consider this as
    the redshift distribution of the source galaxies in our
    simulations.}
  \label{fig4}
\end{figure}

To approximate a reasonable redshift distribution (and hence an
estimate of $f_i(\chi_s)$) for future radio surveys, we make use of
the SKADS simulation of \cite{wilman08}. We select only the
star-forming galaxies from this simulation down to a 1.4 GHz flux
threshold of $0.5$ $\mu$Jy, which is a reasonable approximation to the
detection threshold that might be achieved with the SKA. The
normalized redshift distribution of this subset of star-forming
galaxies is shown in Figure~\ref{fig4}.
The median redshift of our $S_{1.4{\rm GHz}} > 0.5$ $\mu$Jy sample is
$z_m = 2.0$ and there is also clearly a long tail to higher
redshifts. If the SKADS simulation is representative of the true radio
sky then the redshifts of sources in an SKA-like lensing survey will
be significantly larger than those found in planned optical
lensing surveys such as the Dark Energy Survey and EUCLID for which
the median source redshifts will be $\sim 1$. This would obviously be
good news for radio lensing as the signal would be larger on average,
and therefore easier to measure.
 
\begin{figure}
  \centering
  \resizebox{0.45\textwidth}{!}{  
    \rotatebox{-90}{\includegraphics{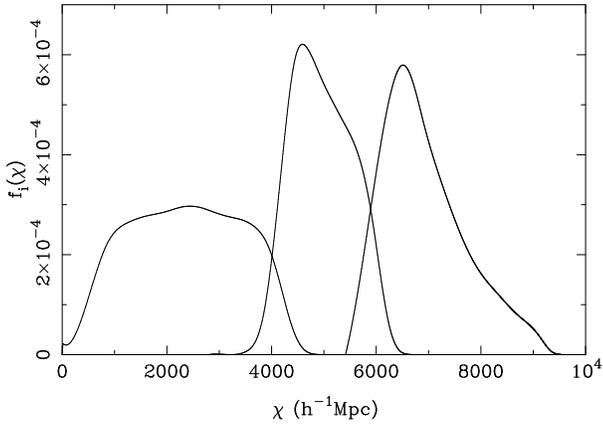}}}
  \caption{Normalized comoving distance distributions for our three
    adopted redshift bins. The overlap in the distributions is due to
    the redshift errors we have introduced. These overlaps in the
    $f(\chi)$ distributions result in non-zero expectation values for
    some inter-bin cross-correlations which would otherwise be
    expected to be zero.}
  \label{fig5}
\end{figure}

For the flux threshold we have applied, the total galaxy number
density is $\sim 18$ arcmin$^{-2}$. Following \cite{blake07}, we note
that in an SKA-like survey, precise redshift information will be
available for a significant proportion of source galaxies via the
detection of their H{\sevensize I} emission. For other galaxies for
which no detection of H{\sevensize I} is obtained, photometric
redshifts could potentially be provided by overlapping multi-band
surveys (e.g.\ from EUCLID). Since our current work is not focused on
the effects of redshift errors, rather than attempting to mimic such a
combination of redshift information, for the purpose of our
simulations, we simply assign random errors to the redshifts of all
sources according to $\sigma_z = (1 + z) \delta_z$ with $\delta_z =
0.05$. The ultimate effect of these redshift errors will be to
introduce a cross-contamination of our chosen redshift bins with
objects from the neighbouring bins. The level of the contamination in
our simulations is likely to be pessimistic for an SKA survey since we
have not considered the precise redshift information coming from the
H{\sevensize I} detections. Taking the redshift distribution of the
star-forming galaxies from the SKADS simulation and folding in the
redshift errors, the normalized selection functions for our three
chosen redshift bins are shown in Fig.~\ref{fig5}.

\subsection{Model for the intrinsic alignment}
\label{sec:ia_model}
To complete the input for our simulations, we require a model for the
IA signal itself. Many authors have attempted to
constrain the IA signal through theory \citep{crittenden01, catelan01,
mackey02, jing02, hirata04}, through observations \citep{brown02, heymans04,
mandelbaum06, mandelbaum09, hirata07, brainerd09} and by measuring the
signal from numerical simulations \citep{heavens00,
  croft00, heymans06}. Notwithstanding these efforts, our understanding of the
effect is currently rather poor, mostly because the underlying physics
of galaxy formation is complicated by gas dynamics, galaxy biasing and
the non-linear evolution of the matter field on small scales. 

One reasonably well motivated theory is the linear alignment model of
\cite{catelan01}, which has subsequently been used to model the IA
signal by a number of authors \citep{hirata04, bridle07}. This model
does not attempt to account for the non-linear evolution of the
density field and so \cite{bridle07} introduced the `non-linear linear
alignment model' whereby they replaced the linear matter power
spectrum with its non-linear counterpart in the linear alignment
model's equations. Although it has effectively no physical motivation,
this non-linear alignment model is probably as good as any other model
for the IA signal available at present. Indeed, \cite{schneider09}
found that a model for the IA signal based on the halo model predicts
a qualitatively similar form for the IA signal. We will adopt the
non-linear alignment model for our simulations but we note again that,
since our technique is model-independent, it would work just as
well in the presence of any other form of IA contamination.

We use a slightly modified version of the non-linear alignment model,
simplified as in \cite{bridle07}. In this model, the IA power
spectrum is simply related to the matter power spectrum via 
\be
P_{\gamia}(k) = \frac{C_1^2 \bar{\rho}^2}{\bar{D}^2} P_{\delta}(k),
\label{eqn:ia_3d_power}
\ee
where $\bar{D}(z) \equiv (1 + z) D(z)$ is the growth factor normalized
to unity at the present day and $\bar{\rho}$ is the mean matter
density of the universe. The constant $C_1$ is a normalization
constant which \cite{hirata04} and subsequent authors have matched to
the amplitude of the II signal measured in
SuperCOSMOS \citep{brown02}, yielding $C_1 = 5 \times 10^{-14} (h^2
M_{\odot} / {\rm Mpc}^{-3})^{-2}$. This amplitude is also consistent
with constraints on the II and GI signals obtained from the SDSS
\citep{mandelbaum06}. The cross power between the matter and
IA fields is given by 
\be
P_{\dgamia}(k) = - \frac{C_1 \bar{\rho}}{\bar{D}} P_{\delta}(k).
\label{eqn:ia_matter_3d_xpower}
\ee
Within this model, we have $P_{\dgamia}(k) = -\sqrt{P_{\gamia}(k)
  P_{\delta}(k)}$. That is, the matter and IA fields are 100\%
anti-correlated. We will see in the next section that this property
presents significant difficulties when it comes to realizing
correlated 2D fields with the correct statistical properties. However,
even before we consider these practical issues, heuristically, one can
imagine that a perfect anti-correlation between the IA and matter
fields is highly unlikely in reality due to the complicated nature of
the galaxy formation process, and hence of the generation of the
intrinsic correlations. We therefore slightly modify the non-linear
alignment model to include a correlation coefficient, $\rho_c$ in the
cross-power spectrum expression:
\be 
P_{\dgamia}(k) =
\rho_c \frac{C_1 \bar{\rho}}{\bar{D}} P_{\delta}(k).
\label{eqn:ia_matter_3d_xpower_corr}
\ee
To retain the connection between the simple physical picture of the
linear alignment model \citep{catelan01, hirata04}, we enforce the
condition, $\rho_c < 0$ so that the matter and IA
fields are (partially) anti-correlated and thus result in a negative
contribution to the lensing power spectra
(c.f.~equation~\ref{eqn:cl_gi}). 

\begin{figure*}
  \centering
  \resizebox{0.80\textwidth}{!}{  
    \rotatebox{-90}{\includegraphics{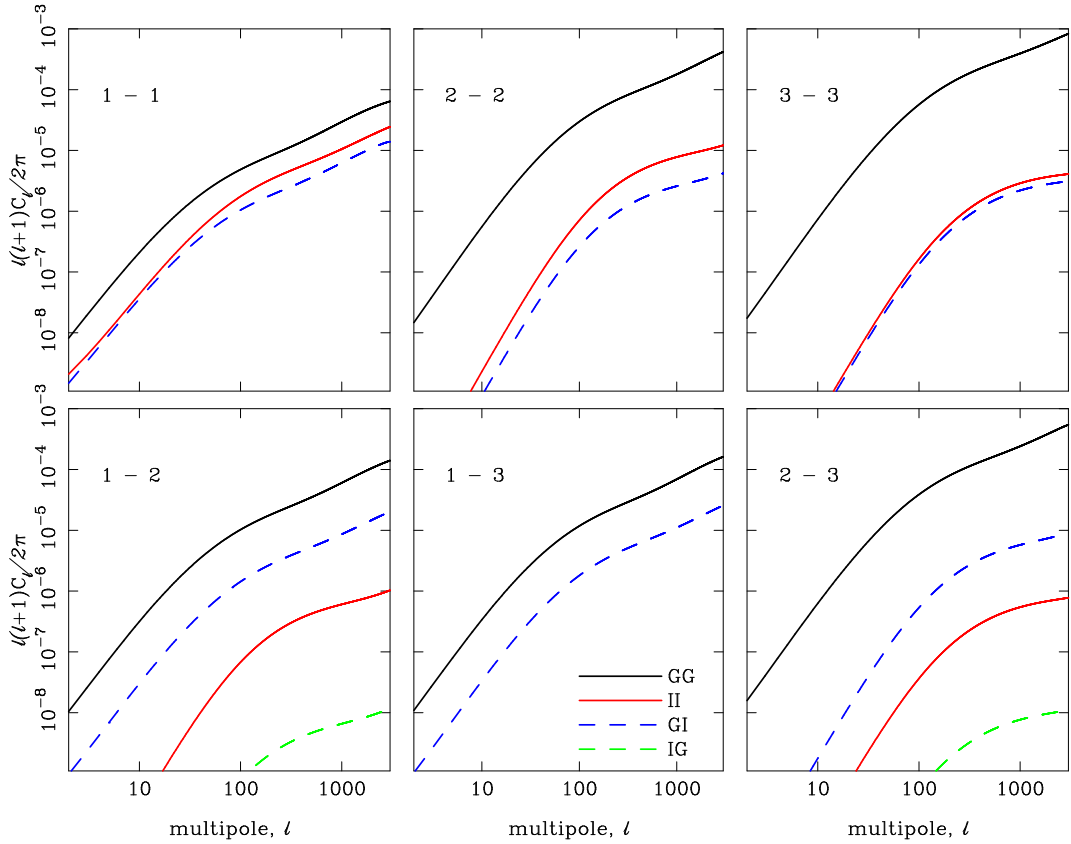}}}
  \caption{Input power spectra used for the simulations for the case
    where we have normalized the IA signal to five times the amplitude
    observed in SuperCOSMOS. The title on each panel indicates which
    two bins are being correlated. E.g.\ `1 - 2' means we’re
    correlating fields in bin 1 with fields in bin 2. Black lines show
    the lensing signal. Red lines show II signals. The broken blue
    (and green) lines show the GI interference terms between lensing
    and IA. Unbroken curves are positive signals whereas the signal
    shown as broken curves are negative. Note the difference between a
    foreground intrinsic-background shear correlation (shown as the blue
    lines) and a background intrinsic-foreground shear correlation (shown
    as the green lines). In the absence of redshift errors, these latter
    types of correlations, and correlations between the intrinsic
    fields in different redshift bins, would be expected to
    vanish. For example, there are no II or IG signals in the `1 - 3'
    panel since the redshift distributions for these bins do not
    overlap (c.f.\ Figure~\ref{fig5}).}
  \label{fig6}
\end{figure*}

We have conducted our analysis for two choices of the model
parameters, $C_1$ and $\rho_c$. For our main analysis, we have used
the SuperCOSMOS normalization and a correlation coefficient of $\rho_c
= -0.2$. In order to demonstrate our technique on a very significant
level of IA contamination, we have also considered a normalization of
five times the SuperCOSMOS value, again with a correlation coefficient
of $\rho_c = -0.2$. Our reason for keeping the correlation at the
apparently low value of 20\% will be made clear in the next
section. Fig.~\ref{fig6} shows the lensing power spectra, the IA power
spectra and the cross-power spectra, calculated using
equations~(\ref{eqn:cl_gg})--(\ref{eqn:cl_gi}), for the case where the
IA normalization is five times the SuperCOSMOS value.

\subsection{Simulating correlated weak lensing and intrinsic alignment fields}
\label{sec:corr_simsl}
From a set of power spectra such as those shown in Fig.~\ref{fig6}, we
wish to generate six correlated fields, one lensing shear field and one
intrinsic shear field in each of our three redshift bins. We will
refer to these fields as G1, G2, and G3 for the lensing shear fields
in redshift bins 1--3 and I1, I2, and I3 for the intrinsic fields in
bins 1--3. We will work on the spherical sky and therefore use the
(spin) spherical harmonic basis. 

To create Gaussian realizations of all six fields with the correct
correlations, at each multipole, $\ell$, we construct the symmetric $6
\times 6$ power spectrum matrix, $C^{xy}_{\ell}$ where $\{x, y\} =
\{G1, G2, G3, I1, I2, I3\}$ and the entries of the matrix are the
power spectra shown in Fig.~\ref{fig6}. Taking the Cholesky
decomposition of this matrix at each multipole, $L^{xy}_\ell$, defined
by \be C^{xy}_{\ell} = \sum_z L^{xz}_\ell L^{yz}_\ell,
\label{eqn:cholesky_dcmp}
\ee
then we can generate random realizations of the spin-2 spherical
harmonic coefficients of each field as
\ba
a^x_{\ell 0} &=& \sum_y L^{xy}_\ell G^y_{\ell 0}, \nn \\
a^x_{\ell m} &=& \sqrt{\frac{1}{2}} \sum_y L^{xy}_\ell G^y_{\ell m},
\label{eqn:alm_sim_fields}
\ea 
where $G^x_{\ell m}$ is an array of unit-norm complex Gaussian random
deviates. The resulting fields transformed to real space via a spin-2
transform will then possess the desired correlations between the
fields. Note that the harmonic modes of
equation~(\ref{eqn:alm_sim_fields}) are the even-parity $E$-modes ---
in addition to assuming that the lensing signal is pure $E$-mode, we
also adopt a pure $E$-mode signal for the IA field, as would be
expected in the linear alignment model
\citep{catelan01}. Consequently, in all of our simulation work we set
the odd-parity $B$-mode component to zero.

In order for the Cholesky decomposition of
equation~(\ref{eqn:cholesky_dcmp}) to be well-defined, the power
spectrum matrix must be positive-definite. More fundamentally, since
the power spectrum matrix is simply the covariance matrix of the
(assumed Gaussian) fields, then it is a requirement that this matrix
be positive semi-definite. If it is not positive semi-definite, then
it is not a valid covariance matrix.

We find that when we use the unmodified non-linear (or linear)
alignment IA model (equations~\ref{eqn:ia_3d_power} and
\ref{eqn:ia_matter_3d_xpower}), non-positive definite power spectrum
matrices result. In fact, during our analysis we have found that for
some configurations of redshift bins and IA signal strength, the
necessary constraint, $|C_\ell^{XY}| \leq \sqrt{C^{XX}_\ell
  C^{YY}_\ell}$ is not satisfied or, in other words, that the two
fields $X$ and $Y$ are more than 100\% (anti-) correlated. These
effects can be traced back to the (arguably unrealistic) 100\%
anti-correlation between the matter and IA fields which these models
assume and is the reason why we have introduced a correlation
coefficient, $|\rho_c| \neq 1$ to describe the strength of the
correlation between the matter and IA fields. The inconsistency may
also be partially due to our approximating the full 3D lensing and IA
fields as correlated 2D fields in three redshift bins. A detailed
investigation into the degree of correlation one might reasonably
expect for the IA and matter fields is beyond the scope of this
paper. For the purposes of our simulations, we simply set the matter
and IA fields to be anti-correlated at the 20\% level which, for our
configuration of redshift bins is the largest degree of correlation
which results in a positive definite power spectrum matrix at all
multipoles.

Using our model for the lensing shear and IA signals, we generate the
$E$-mode spin-spherical harmonics of all fields using
equations~(\ref{eqn:cholesky_dcmp}) and (\ref{eqn:alm_sim_fields}) up
to a maximum multipole of $\ell_{max} = 2048$. This value was chosen
in order to avoid the strongly non-linear regime as well as for
computational ease. We set the $B$-modes of all fields to zero and use
fast spin-2 spherical harmonic transform routines from the
\healpix\footnote{see http://healpix.jpl.nasa.gov/index.shtml and
  \cite{gorski05}} package to transform these to real space. The
resulting fields are pixelised with a resolution of $\sim 3.4$ arcmin
(\healpix\, resolution parameter, $N_{\rm side} = 1024$).

\subsection{Generating the observable fields}
To simulate observable quantities, in each pixel of each redshift bin,
we generate the `observed' ellipticity of a finite number of
galaxies as 
\be \vepsilon^{\rm obs} = \vgamma^G + \vgamma^I + \vepsilon^{\rm rand},
\label{eqn:ellip_sim}
\ee
where $\vgamma^G$ and $\vgamma^I$ are the lensing and IA fields
generated using the procedure described in the previous section and
are the same for all galaxies within a pixel. $\vepsilon^{\rm rand}$ is
the random shape noise associated with the intrinsic dispersion in
galaxy ellipticities and is different for each galaxy in a pixel. For
the random shape noise, we assume a Gaussian distribution with an r.m.s.
ellipticity of $\vepsilon_{\rm rms} = 0.3$.  We use a galaxy number
density of 6 arcmin$^{-2}$ for each of our three redshift bins which
were chosen to be equally populated (see
Section~\ref{sec:cosmo_model}). The total number density of galaxies
for which we can measure shape information in our simulations is
therefore 18 arcmin$^{-2}$.

In addition to observed ellipticities for each galaxy, for our
analysis including polarization information, we simulate the
orientation of the observed polarized emission for a subset of the
galaxies according to 
\be 
\alpha^{\rm obs} = \alpha^{\rm int} + \alpha^{\rm rand},
\label{eqn:pa_sim}
\ee
where $\alpha^{\rm int}$ is the intrinsic position angle of the galaxy,
\be
\alpha^{\rm int} = \frac{1}{2} \tan^{-1} \left(\frac{\epsilon^{\rm int}_2 +
  \epsilon^{\rm rand}_2}{\epsilon^{\rm int}_1 + \epsilon^{\rm rand}_1} \right).
\label{eqn:pa_int}
\ee
Following the discussion in Section~\ref{sec:pol_morph}, for a given
instrument sensitivity, there is clearly a trade off in how one
defines the detection threshold for polarization --- as one increases
the threshold, the number of sources for which we have intrinsic
position angle information will decrease, but the noise on those
position angle estimates ($\alpha_{\rm rms}$) will also decrease. This
behaviour will hold until the limiting irreducible astrophysical
scatter in the polarization orientation--intrinsic position angle
relationship is reached. In a real analysis, one could envisage
retaining all the galaxies in the sample for the polarization analysis
and choosing the weights of equations~(\ref{eqn:Amatrix_def}) and
(\ref{eqn:bvec_def}) based on the signal-to-noise of the polarization
measurements. For our simulations, we take a simpler route and assume
that we can measure the orientation of the polarized emission for 10\%
of the galaxies in the sample and that the resulting estimates of the
intrinsic position angles are subject to a combined measurement and
astrophysical scatter of $\alpha_{\rm rms} = 5^\circ$. We use this
value to add Gaussian noise to our intrinsic position angle estimates
via $\alpha^{\rm rand}$ in equation~(\ref{eqn:pa_sim}) and we subsequently
assign uniform weights to all galaxies in our polarization sub-sample.

In summary, our simulated observations consist of the noisy
ellipticities of galaxies containing a shear and IA signal as
constructed via equation~(\ref{eqn:ellip_sim}) and the noisy estimates
of the intrinsic position angles of 10\% of the galaxies as
constructed via equation~(\ref{eqn:pa_sim}).

\section{Analysis}
\label{sec:analysis}
In order to test the effectiveness of the technique presented in
Section~\ref{sec:estimator}, we will reconstruct shear maps and
estimate the various power spectra from our simulated observations
using both the standard lensing estimator and using our new technique
making use of polarization information. 

\subsection{Shear maps}
\label{sec:maps}
For the standard estimator, for each of our three redshift bins, we
reconstruct the shear signal in each of the 3.4 arcmin pixels using
equation~(\ref{eqn:standard_est}). This estimator uses the full number
density of simulated galaxies, i.e.\ 6 galaxies arcmin$^{-2}$ in each
redshift bin. Of course, the reconstructed shear fields will be
contaminated by the IA signal. For the new estimator, we apply
equations~(\ref{eqn:new_est})--(\ref{eqn:bvec_def}) to the 10\% of
galaxies in each pixel for which we have simulated intrinsic position
angle estimates. The number density used for the new estimator is
therefore 0.6 galaxies arcmin$^{-2}$ per redshift bin. As described in
the previous section, we weight each galaxy equally for this analysis.

\begin{figure*}
  \centering
  \resizebox{0.98\textwidth}{!}{  
    \rotatebox{0}{\includegraphics{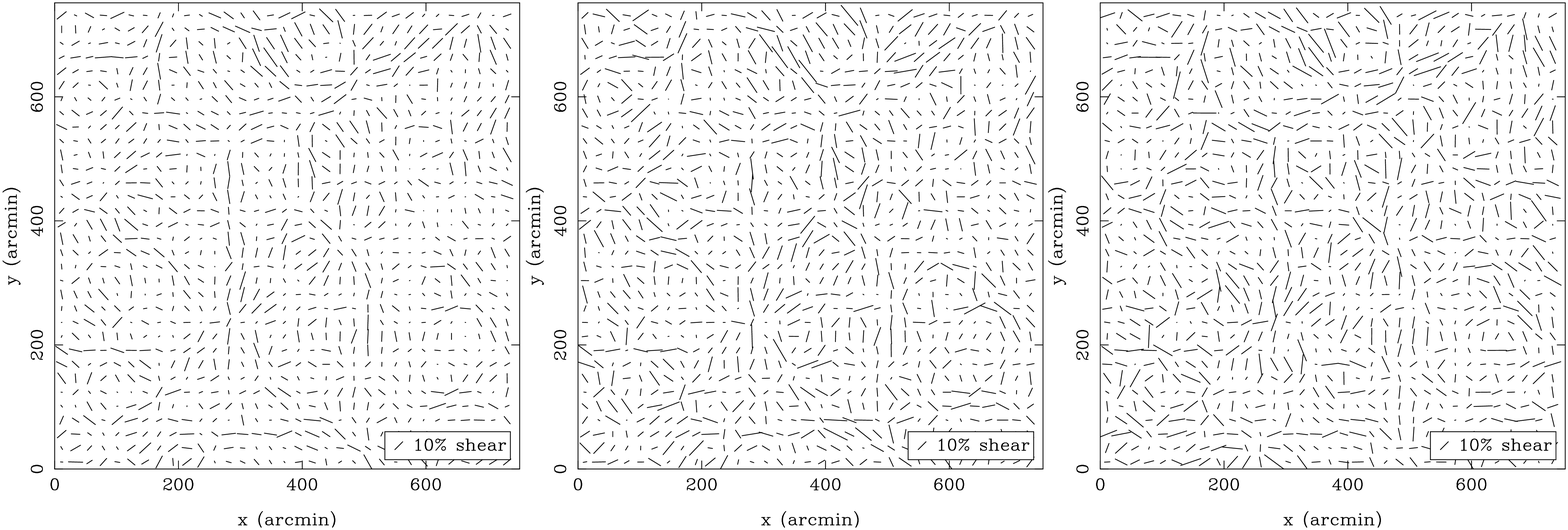}}}
  \caption{Example of the reconstruction of the shear signal for our
  highest redshift bin. The input signal is shown in the left
  panel. The central panel shows the reconstructed signal using the
  standard estimator using a galaxy number density of $6$
  arcmin$^{-2}$. The right panel shows the reconstruction using the
  new estimator with a factor ten less galaxies and for a scatter in
  the intrinsic position angle estimates of 5 degs. For this plot, we
  have re-sampled the shear signals from the spherical sky onto a
  $12.5 \times 12.5$ deg$^2$ Cartesian grid with pixel size 4.5
  arcmin.}
  \label{fig7}
\end{figure*}

Fig.~\ref{fig7} shows an example of the reconstruction of the lensing
signal, over a $\sim 150$ deg$^2$ region in our highest redshift bin,
projected from the spherical sky maps onto a Cartesian grid. In this
redshift bin, the IA signal is negligible compared to the lensing
signal and so the reconstructed maps can be directly compared to the
input signal. It is clear from this figure that, for the parameters
which we have adopted, the new estimator recovers the input signal
with a similar precision to that achieved with the standard estimator.

\subsection{Power spectrum estimation}
\label{sec:cl_estimation}
To estimate the power spectra from the reconstructed shear maps, we use
a standard pseudo-$C_\ell$ approach \citep{hivon02, brown05}. These
fast power spectrum techniques have been widely used to analyze large
CMB temperature and polarization datasets. In principle, the
extension to lensing is straight-forward via the transformation $\{I,
Q, U, E, B\} \rightarrow \{\mu, \gamma_1, \gamma_2, \kappa, \beta\}$
where $I, Q$ and $U$ are the Stokes parameters of the CMB field and
$\mu, \gamma_1$ and $\gamma_2$ are the magnification and shear
components of the lensing fields. $E$ and $\kappa$ denote the
even parity $E$-modes of the CMB polarization and lensing fields
respectively (where for the latter, we can further identify the
$E$-modes as the lensing convergence field). $B$ and $\beta$
are the odd-parity $B$-modes of the CMB polarization and lensing
fields respectively. For lensing, we do not expect a significant
cosmological signal in $\beta$ and so we ignore the $B$-modes in the
analysis which follows.

Although we do not address them in this paper, we note that, in
practice, there are a number of real-world issues which make the
extension of pseudo-$C_\ell$ power spectrum estimators to weak lensing
more problematic. Most importantly, in contrast to the simple
apodizing masks usually adopted in CMB analyses, lensing analyses
typically involve complicated and highly irregular masking of the data
to remove diffraction spikes from bright stars and other localized
contamination. Defining the optimal mask for such complicated survey
geometries is not trivial (although see \citealt{hikage10} for a
recent investigation into some of these issues). In the simulations
which follow, while acknowledging that a full-sky lensing survey is
unrealistic, we avoid all of these issues, including any $E
\leftrightarrow B$ mixing effects by working on the full (and
complete) sky.

Our estimated shear maps are first transformed to spherical harmonic
space using the \healpix\, spin-2 transform routines. We discard the
$B$-modes (which are consistent with noise) and by taking averages of
the harmonic $E$-modes of the maps, one can estimate the
pseudo-$C_\ell$ power spectra, $\tilde{C}^{XY}_\ell$ via 
\be
\tilde{C}^{XY}_\ell = \frac{1}{2\ell + 1} \sum^{m = +\ell}_{m=-\ell}
a^X_{\lm} {a^Y_{\lm}}^*,
\label{eqn:pcl_est}
\ee
where $X$ and $Y$ denote the two fields being correlated and the
asterix denotes complex conjugation. For noise free observations over
the full sky, the power spectra estimated via
equation~(\ref{eqn:pcl_est}) should be unbiased (apart from the effect
of the map pixelization which we correct for). In the presence of
noise (but still on the full sky), the expectation value of this
estimator is usually taken to be
\be
\lgl \tilde{C}^{XY}_\ell \rgl = C^{X_sY_s}_\ell + C^{X_nY_n}_\ell, 
\label{eqn:pcl_expval}
\ee
where $C^{X_sY_s}_\ell$ is the true cosmological signal and $C^{X_nY_n}_\ell$
is the power spectrum of the noise. A correction is therefore usually
applied by measuring the noise bias, $C^{X_nY_n}_\ell$ from a suite of
noise-only simulations and subsequently subtracting this from the
measured spectra,
\be
\hat{C}^{XY}_\ell = \tilde{C}^{XY} - \lgl C^{X_nY_n}_\ell \rgl_{\rm mc},
\label{eqn:pcl_debias}
\ee 
where the angled brackets denote an average over Monte-Carlo
simulations containing only noise. For the standard estimator, these
noise-only simulations consist of just the random noise in the galaxy
ellipticities. For the new estimator, they consist of both the random
ellipticity noise and the random noise in the intrinsic position angle
estimates. 

We use power spectra estimated using equation~(\ref{eqn:pcl_debias})
as our main diagnostic for assessing the impact of IA contamination
and the effectiveness of our new lensing estimator in removing
it. Once estimated, we bin the $C_\ell$'s into 32 equal-width flat
bandpowers (denoted $P_b$) spanning our entire multipole range ($2 <
\ell < 2048$). Since the lensing (and IA) power spectra do not exhibit
any significant features over $\ell$-ranges comparable to our bin-size
($\Delta\ell = 64$), we effectively lose no information by performing
this binning but the plots are much less cluttered and easier to
interpret.

Finally, the covariance matrix of the bandpowers can be estimated from 
the scatter among a suite of (in our case, 200) Monte-Carlo simulations: 
\be
\lgl \Delta P^X_b \Delta P^Y_{b'}\rgl = 
\lgl(P^X_b - \overline{P}^X_b)(P^Y_{b'} - \overline{P}^Y_{b'})
\rgl_{\rm mc} \,,
\label{eqn:pcl_covar}
\ee
where the overline denotes the mean over all simulations. 

\subsection{Lensing power spectra results}
\label{sec:lensing_results}
\begin{figure*}
  \centering
  \resizebox{0.90\textwidth}{!}{  
    \rotatebox{-90}{\includegraphics{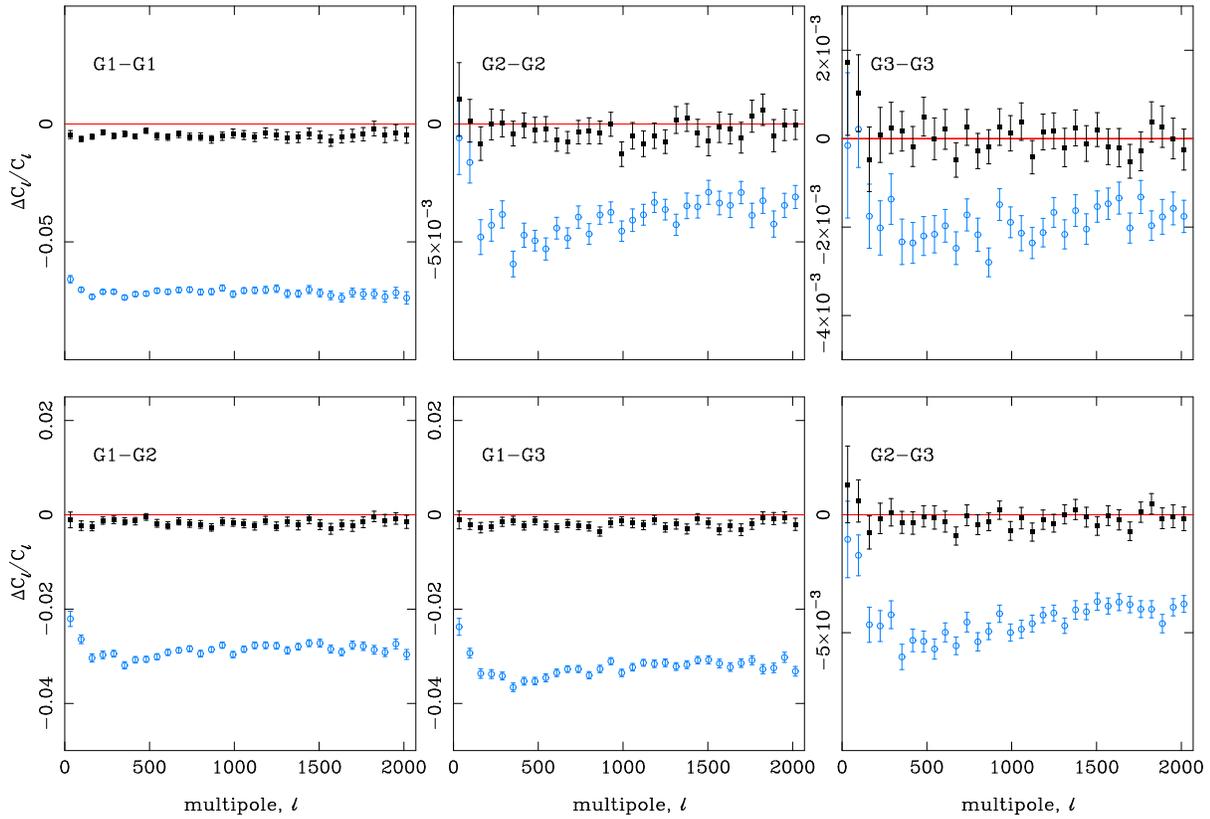}}}
  \caption{Fractional bias in the reconstruction of the lensing power
  spectra for the standard estimator (light blue points) and for the
  new estimator including polarization information (black
  points). These results are for the simulations where the amplitude
  of the IA contamination was set to the level seen in
  SuperCOSMOS. Once again, a factor ten less galaxies was used for the
  new estimator and we assumed an intrinsic position angle scatter of 5
  degs. For all six spectra, the average reduction in the bias is more
  than an order of magnitude.}
  \label{fig8} 
\end{figure*}

The power spectrum results for the case where the IA signal was
normalized to the amplitude seen in SuperCOSMOS are shown in
Fig.~\ref{fig8}. 
For this level of IA contamination, the biases in the
power spectra are rather small compared to the range in amplitude of
the lensing spectra over our full multipole range. In Fig.~\ref{fig8},
instead of plotting the power spectra themselves, we therefore plot
the fractional bias, $\Delta C_\ell / C^{\rm input}_\ell$ where $\Delta
C_\ell$ is the difference between the mean recovered spectra and the
input model. Comparing the results from the standard and new
estimators we see that our technique has successfully reduced the IA
bias in each spectrum by over an order of magnitude on all scales. The
mean fractional bias across all multipoles for each spectrum for the
standard and new estimators are presented in Table~\ref{tab1}. There
is a small residual bias in the spectra recovered using the new
estimator but, as we shall see in Section~\ref{sec:params_results}, this has a
negligible impact on our inferences regarding the underlying
cosmology. (Note that the error-bars presented in Fig.~\ref{fig8} are
the errors on the mean recovered spectra, not those for a single
realization which would be $\sqrt{200}$ larger.)

In Table~\ref{tab1}, we also present the fractional biases in the
recovered power spectra for our simulations which included an IA
signal normalized to five times the signal seen in SuperCOSMOS. For
the majority of the spectra, using the new estimator, the bias is once
again reduced by a factor of ten or more. The exception is for the
auto-power measured in our lowest redshift bin where the bias is
reduced by only a factor of $\sim 3$. The reason for this relatively
poor performance is related to the apparent anomaly that the bias in
the standard estimator decreases when we increase the amplitude of the
IA contamination by a factor of five. This effect is, in fact, due to
a fortuitous part-cancellation of the positive II and negative GI
signals within our lowest redshift bin when using the standard
estimator. This cancellation is stronger in our simulations for the
higher amplitude IA signal. Obviously, since such cancellations are in
no way guaranteed and are highly dependent on both the details of the
IA signals and the choice of redshift binning, we argue that the
result for the $G1-G1$ bias in the standard case for the larger IA
amplitude is misleadingly low. Furthermore, comparing between the
numbers for the two sets of simulations, we see that for the new
estimator, the residual bias increases by a factor of $\sim 5$ when we
increase the IA amplitude by a factor of 5, suggesting that the degree
to which the IA signal is removed using the new estimator is
independent of the IA model.

\begin{table}
\caption{Mean fractional bias across all multipoles in the recovered lensing
  power spectra for the standard and new shear estimators. The first
  set of numbers report the biases in simulations where the IA signal
  is normalized to the amplitude seen in SuperCOSMOS. The second set
  of numbers are for an IA signal five times larger.}
\begin{center}
\begin{tabular}{c|r|r}
Spectrum & Standard estimator & New estimator \\
\hline
1 $\times$ SuperCOSMOS: & & \\
$G1-G1$ & $-7.14 \times 10^{-2}$ & $-4.85 \times 10^{-3}$ \\
$G2-G2$ & $-3.84 \times 10^{-3}$ & $-2.55 \times 10^{-4}$ \\
$G3-G3$ & $-1.80 \times 10^{-3}$ & $ 8.38 \times 10^{-5}$ \\
$G1-G2$ & $-2.85 \times 10^{-2}$ & $-1.67 \times 10^{-3}$ \\
$G1-G3$ & $-3.23 \times 10^{-2}$ & $-1.95 \times 10^{-3}$ \\
$G2-G3$ & $-4.36 \times 10^{-3}$ & $-1.45 \times 10^{-4}$ \\
        &                       &                        \\  
5 $\times$ SuperCOSMOS: & & \\
$G1-G1$ & $-6.34 \times 10^{-2}$ & $-2.26 \times 10^{-2}$ \\
$G2-G2$ & $ 1.27 \times 10^{-2}$ & $-1.30 \times 10^{-3}$ \\
$G3-G3$ & $-3.31 \times 10^{-3}$ & $-5.40 \times 10^{-4}$ \\
$G1-G2$ & $-1.34 \times 10^{-1}$ & $-8.43 \times 10^{-3}$ \\
$G1-G3$ & $-1.60 \times 10^{-1}$ & $-9.40 \times 10^{-3}$ \\
$G2-G3$ & $-1.97 \times 10^{-2}$ & $-1.33 \times 10^{-3}$ \\
\hline
\end{tabular}
\end{center}
\label{tab1}
\end{table}

\subsection{Intrinsic alignment reconstruction}
\label{sec:ia_results}
In addition to the lensing signal, a combination of the standard and
new shear estimators gives us an estimate of the IA signal itself via
equation~(\ref{eqn:ia_est}). In the limit of perfect intrinsic
position angle measurements, our new lensing estimator would be
entirely shot noise free. However, this is not the case for estimates
of the IA signal which are subject to the shot noise of the standard
estimator. This noise will be increased further due to the errors
in the position angle estimates. Since this noise is very much larger
than the IA signal, our reconstructed IA maps are noise-dominated and
are not particularly informative. However, we can still usefully
constrain the IA signal in the power spectrum.

\begin{figure*}
  \centering
  \resizebox{0.95\textwidth}{!}{  
    \rotatebox{0}{\includegraphics{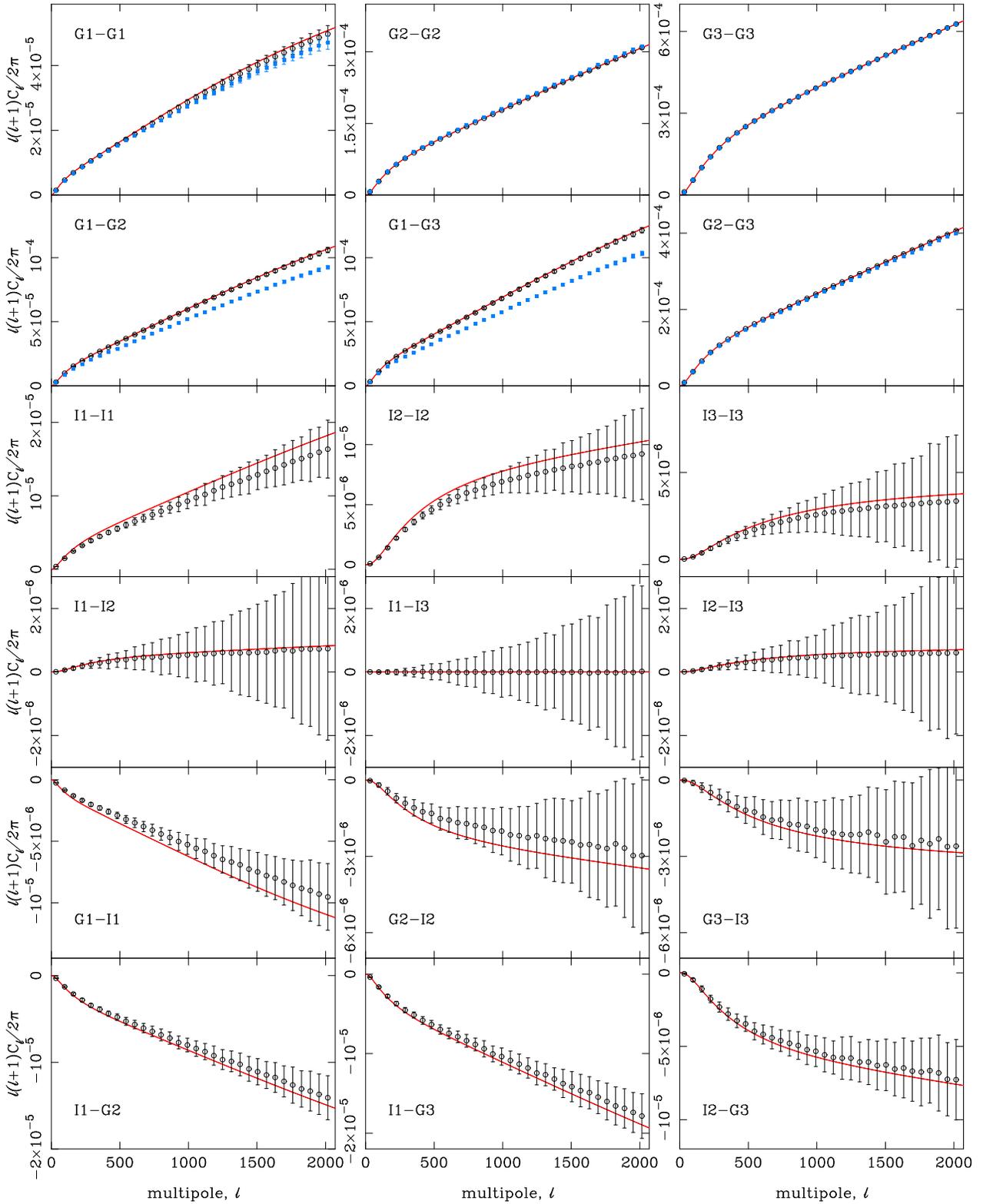}}}
  \vspace{0.5cm}
  \caption{Reconstruction of the lensing, IA and cross-correlation
    signals for an IA amplitude of five times the SuperCOSMOS
    level. The input model spectra are shown as the smooth curves.
    For the lensing spectra (top two rows), the standard
    estimator measurements are shown as the light blue points and
    the spectra measured using the new estimator are shown as the
    black points (for the higher redshift correlations, where the
    lensing signal dominates, the two sets of points are
    indistinguishable on this plot). The other panels show the
    reconstruction of the IA spectra and lensing-IA cross correlations
    using a combination of the standard and new estimators. The points
    show the mean recovered spectra averaged over all simulations and
    the error-bars are those appropriate for a single realization. We
    omit the reconstruction of the foreground shear --- background
    intrinsic cross-correlations as these signals are very small and
    effectively unconstrained in our simulations (c.f.\ the dashed
    green curves in Fig.~\ref{fig6}).}
  \label{fig9} 
\end{figure*}

Fig.~\ref{fig9} shows the simultaneous reconstruction of the lensing
spectra, the IA spectra and the lensing--IA cross-correlations for the
case where we've normalized to five times the SuperCOSMOS level. All
of these spectra have been estimated from the reconstructed lensing
and IA maps using the normal pseudo-$C_\ell$ estimator of
equation~(\ref{eqn:pcl_debias}). For the lensing spectra, we also plot
the recovered signal from the standard lensing estimator for
comparison. Examining the IA spectra and the lensing-IA
cross-correlations, we see a biased reconstruction. The source of
this bias is the same as for the biases already seen in the recovered
lensing signal --- effectively, it is an additional `noise bias'
caused by the presence of the IA signal itself and the
noise in our intrinsic position angle estimates. Despite the bias,
we see that the general form of the IA signals and the lensing-IA
cross-correlations is recovered rather well --- the residual
fractional bias is fairly constant across multipoles for all
spectra. Such reconstructions of the IA signal with future datasets
could potentially provide much needed observational constraints on
theoretical models of the IA signal. 

\subsection{Parameter constraints}
\label{sec:params_results}
We consider the power spectrum results of
Section~\ref{sec:lensing_results} as the main diagnostic of the
performance of the new estimator in removing a potential IA
contaminant. It is also interesting, however, to examine the impact of
our new technique on the reduction of biases in cosmological parameter
constraints. As already pointed out, our simulations are clearly
unrealistic in that they assume full and complete sky coverage. In
addition, to conduct our analysis, we have also had to adopt values
for some parameters which are currently very uncertain. In addition to
the amplitude and form of the IA signal, these include the typical
fractional polarization of faint star-forming galaxies and the
strength of the correlation between the orientation of the polarized
emission and intrinsic morphologies of these galaxies. Consequently,
the constraints on parameters obtained from these simulations will very
likely be over-optimistic. We therefore emphasize that the purpose of
this section is not to make parameter forecasts for lensing surveys
with the SKA or any other instrument. Rather, our purpose here is simply
to demonstrate that our proposed technique could potentially have a
large impact in terms of minimizing biases in cosmological constraints
from future radio lensing surveys.

We adopt a very simple approach and perform a standard grid-based
likelihood analysis in three cosmological parameters which cosmic
shear measurements can strongly constrain --- the matter density
($\Omega_m$), the power spectrum normalization ($\sigma_8$) and the
(assumed constant) equation of state of the dark energy ($w$). We
enforce a flat Universe ($\Omega_\Lambda = 1 - \Omega_m$) and keep all
other parameters fixed at their fiducial values (see
Section~\ref{sec:cosmo_model}). As our `data', we use only the six
lensing power spectra constrained in Section~\ref{sec:lensing_results}
and we perform the analysis for the mean power spectra estimated using
both the standard and new estimators. Ordering these spectra into a $6
\times N_{\rm band}$ data vector ($\vd$), we calculate the likelihood
at each point in parameter space according to
\be -2 \ln\mathcal{L} = (\vd - \vd^{\rm th}) \mC^{-1} (\vd - \vd^{\rm th})^T, 
\ee 
where $\vd^{\rm th}$ are the model power spectra binned in the same way as
our simulated data. The covariance matrix, $\mC$ is calculated from
the simulations according to equation~(\ref{eqn:pcl_covar}) but we
retain only the diagonal elements and all same-bin inter-spectra
covariances. The remainder of the covariance matrix is set to zero
since these elements are non-zero only because of numerical noise from
the finite number of simulations. (Since our simulations are on the
full sky, measurements at different multipoles will be uncorrelated.)

\begin{figure*}
  \centering
  \resizebox{0.90\textwidth}{!}{  
    \rotatebox{-90}{\includegraphics{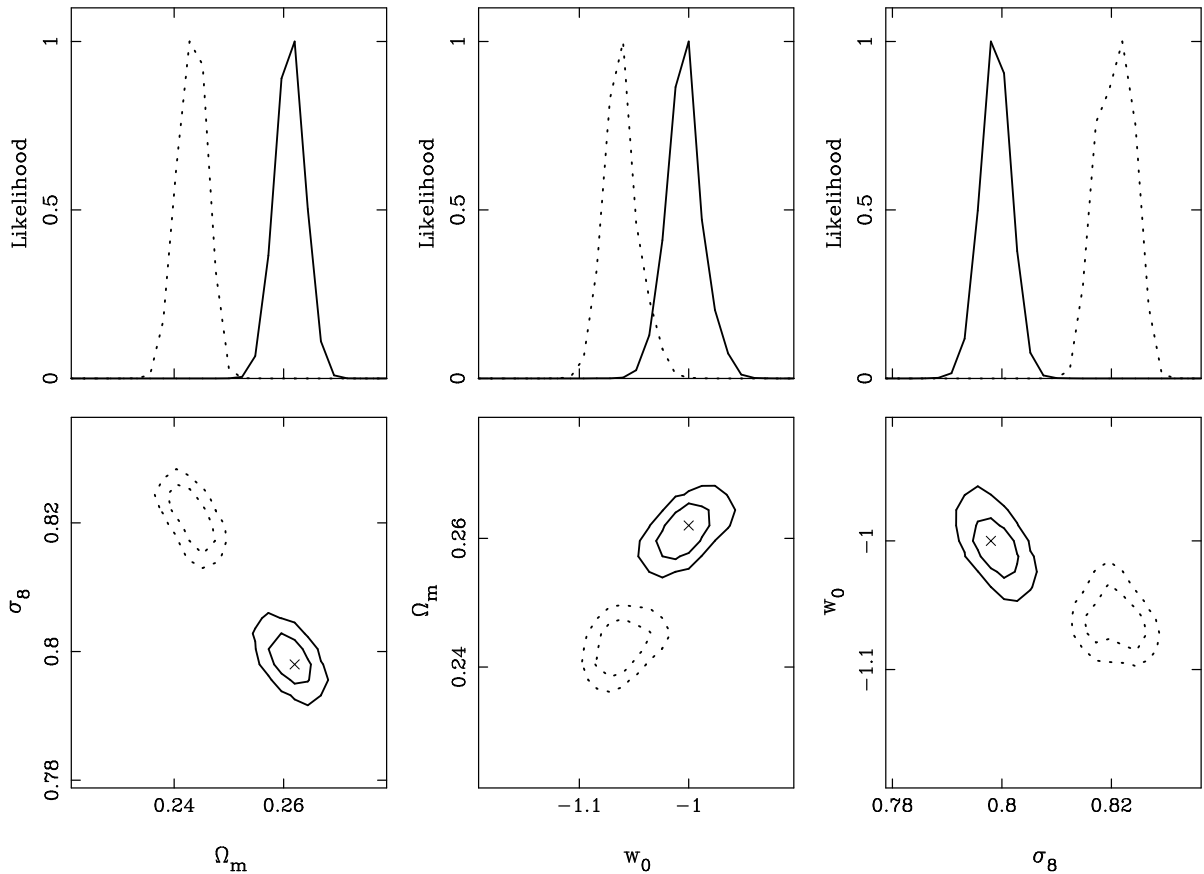}}}
  \caption{Marginalized 1D and 2D parameter constraints recovered from
    simulations where the IA contamination is of order that seen in
    SuperCOSMOS. For our maximum likelihood analysis, we have only
    varied the three parameters, $\Omega_m$, $w$ and
    $\sigma_8$. Constraints marginalized over other parameters would
    be weaker. The dotted lines show the constraints obtained using
    the standard estimator. The full lines show the constraints using
    our new estimator and a factor ten less galaxies. The input model
    is indicated in the lower panels with a cross. Including the
    polarization reduces parameter biases to negligible levels with
    essentially no increase in errors. For the 2D constraints the
    contours indicate the 68\% and 95\% confidence regions.}
  \label{fig10} 
\end{figure*}

The resulting 1D and 2D parameter constraints are shown in
Figure~\ref{fig10} for the simulations where the IA signal was set to
the level seen in SuperCOSMOS. For the 1D constraints, we have
marginalized over the other two parameters which are varied, and for
the 2D constraints, we have marginalized over the single remaining
parameter. In addition to the caveats we mention above regarding the
unrealistic aspects of our simulations, we note that constraints
marginalized over additional parameters, or those obtained in an
extended parameter space would obviously be weaker. We see that for
all three parameters, using the standard estimator results in very
significant biases in the constraints obtained. Moreover, the
size of the biases seen in Fig.~\ref{fig10} are at least as large as
the $1\sigma$ uncertainties in cosmological parameters expected to be
achieved with the ``Stage IV'' dark energy projects identified by the
Dark Energy Task Force \citep{albrecht06} including the SKA. 

The biases are reduced to negligible levels when we use the power
spectra obtained with the new shear estimator. For the latter case,
the maximum likelihood model corresponds to the input model. Note that
the size of the errors are not particularly informative (due to the
caveats we have already mentioned) but the relative size of the errors
for the standard and new shear estimator cases is relevant and is
extremely encouraging. That is, the bias due to the IA contamination
has been removed (essentially completely) with effectively no loss in
cosmological information.

We have also performed the likelihood analysis for the simulations
with the much larger IA normalization. In this case, there are
residual $\sim 1\sigma$ biases in the parameters obtained with the new
estimator but the degree to which the contamination is reduced from
the standard estimator case is similar to that seen in Fig.~\ref{fig10}.

\subsection{A procedure to correct for residual bias}
\label{sec:bias_correction}
Although the results of the previous sections are encouraging, ideally
one would prefer a completely unbiased and model-independent technique
for removing IA signals from lensing surveys. Here we suggest an
iterative technique, which is only mildly model-dependent, to correct
for the residual biases seen in the previous sections. 
 
The expectation values for the pseudo-$C_\ell$ power spectra as written
in equation~(\ref{eqn:pcl_expval}) assume that there is no correlation
between the signal and the noise and this is usually the
case. However, as described in Section~\ref{sec:estimator}, in the
presence of a non-zero IA signal and a non-zero error on our estimates
of the intrinsic position angles, there will be an extra `noise bias'
in our estimates of the shear in each pixel. This bias depends on the
IA signal (see Figure~\ref{fig3}). Although our noise-only
simulations include the noise on the intrinsic position angle
estimates, they do not account for this extra bias since the IA signal
is not present in these simulations. The residual biases in the
power spectra can effectively be eliminated by including an
estimate of the IA signal in the noise-only simulations. In addition,
since non-zero IA signals partly determine the noise in the shear
estimator, we also need to consider correlations between the signal
and the noise in addition to the usual auto-correlations of the noise.

In the most general case, we can write the expectation value of
equation~(\ref{eqn:pcl_est}) as:
\be
\lgl \tilde{C}^{XY}_\ell \rgl = C^{X_sY_s}_\ell + C^{X_nY_n}_\ell +
C^{X_sY_n}_\ell + C^{X_nY_s}_\ell, 
\label{eqn:pcl_expval_ext}
\ee
and one can construct estimators for the various power
spectra, which will always be formally unbiased as
\be
\hat{C}^{XY}_\ell = \tilde{C}^{XY}_\ell - \lgl C^{X_nY_n}_\ell \rgl_{\rm mc} -
\lgl C^{X_s Y_n}_\ell \rgl_{\rm mc} - \lgl C^{X_n Y_s}_\ell \rgl_{\rm mc}.
\label{eqn:pcl_debias_ext}
\ee
Of course, in order to estimate quantities such as $\lgl C^{X_s
  Y_n}_\ell \rgl_{\rm mc}$ from simulations, one requires a model for the
signal component of these simulations.  We have confirmed that using
equation~(\ref{eqn:pcl_debias_ext}) returns completely unbiased results for
all of the power spectra shown in Fig.~\ref{fig9} provided we use the
correct model for the IA signal in the noise-only simulations.  The
power spectrum estimators of Section~\ref{sec:cl_estimation} already
provide us with an initial estimate of the IA signal. An iterative
technique thus presents itself whereby one could fit a theoretical
model (or alternatively use a `model-independent' spline fit) to the
initial power spectrum estimates of the previous sections and then use
the best fitting model as an input IA signal into the noise-only
simulations. Updated estimates of the power spectra could then be
obtained using equation~(\ref{eqn:pcl_debias_ext}) and the process
iterated until subsequent iterations of the estimated power
spectra were deemed consistent. Although we have not tested this
proposed technique, since the recovered IA signal is already very
close to the true model in Fig.~\ref{fig9}, it seems likely that a
single iteration would be sufficient to reduce the residual bias to a
level well below the statistical noise.  

\section{Discussion}
\label{sec:discussion}

Future deep, wide field radio surveys will present a significant
opportunity to measure weak lensing in the radio band with high
precision for the first time. For such precise measurements, the
effects of intrinsic alignments can no longer be ignored. In this
paper, we have presented a technique to mitigate against IA
contaminants in radio weak lensing analyses which makes use of the
fact that the orientation of the polarized emission from background
sources is both unaffected by gravitational lensing, and is related to
the intrinsic morphological orientation of the source.

We have demonstrated the technique on simulated weak lensing skies
including a contaminating IA signal. The results of these tests
suggest that our method has the potential to both reduce shot noise
and to strongly reject IA contamination, in a model-independent way,
with negligible loss in cosmological information. We note that there
are a number of limitations to the simulations presented in this
paper. In particular, we have approximated the lensing shear signal as
being constant for all galaxies within each pixel of each of our three
redshift bins. Of course, in reality, the shear undergone by each
galaxy will depend on both its angular position within the pixel and
on its redshift. This is, however, not an obstacle for our proposed
technique. In the absence of IA signals, it is easy to show that our
new estimator for the shear in each pixel simply returns a weighted
average of the shears sampled at each galaxy position when those
shears are different. This average shear is, of course, the
cosmological observable of interest --- the effects of sub-pixel
gradients in the true shear field and the dependence on redshift of
the average shear field are easily accounted for when generating model
predictions for comparison with observations. In the presence of both
a scatter in the intrinsic position angle estimates and a non-zero IA
signal, there remains a residual bias in the estimator. However, we
have already demonstrated that this bias is small and, if necessary,
could be corrected for.

Although we have presented our technique in the context of weak
lensing in the radio band, an extension to other wavebands, and in
particular to the optical, is in principle possible and would be
highly desirable. In this regard, \cite{audit99} have already
suggested that the orientation of the optical polarized emission from
a galaxy also provides a proxy for the galaxy's intrinsic
orientation. However, the underlying mechanism for the generation of
the polarization is quite different between the radio and optical
bands --- in the former case, the dominant source of polarization is
synchrotron radiation whereas in the latter case, scattering by dust
can also be an appreciable source. In addition, sensitive polarization
measurements are more difficult, from an instrumental point of view,
in the optical band.

Our proposed technique is clearly dependent on the true nature of the
polarized emission from the faint star forming galaxies which will
dominate the number counts of future radio surveys. In particular, the
typical polarization fraction of such galaxies at $\mu$Jy flux
densities and the degree to which the orientation of the polarized
emission is aligned with the morphological orientation are currently
very uncertain. In addition, to apply our analysis to real data, we
will need to correct the observed polarization for the rotation of the
plane of polarization due to intervening cosmological magnetic fields
and the effects of Faraday rotation internal to the sources
themselves. These corrections should be possible by extracting the
rotation measures for the sources using multi-frequency
information. We hope to investigate many of these practical issues and
perform the first application of our proposed technique on powerful
new radio datasets from the imminent e-MERLIN and MeerKAT
instruments in the not-too-distant future.

\section*{Acknowledgments}
We are grateful to Paddy Leahy, Ian Browne, Neal Jackson, Anthony
Challinor and Lindsay King for useful discussions.  The simulation
work described in this paper was carried out on the University of
Cambridge's distributed computing facility, \camgrid. We acknowledge
the use of the \healpix\, \citep{gorski05} package.

\bibliographystyle{mn2e}
\bibliography{ms}

\onecolumn
\appendix
\section{Dispersion of the new estimator}
\label{app:estimator_properties}
If we define for each galaxy within a pixel
\be
{\vn}^\|_i = \left( \begin{array}{c} \cos 2\alpha_i^{\rm int} \\ \sin 2\alpha_i^{\rm int} \end{array} \right) ;
{\vn}^\perp_i = \left( \begin{array}{c} \sin 2\alpha_i^{\rm int} \\ -\cos 2\alpha_i^{\rm int} \end{array} \right), 
\label{eqn:A1}
\ee
then the observed ellipticities can be written as
\be
\vepsilon^{\rm obs}_i = \vepsilon^{\rm int}_i {\vn}^\|_i + \delta \vepsilon_i + \vgamma,
\label{eqn:A2}
\ee
where $\delta \vepsilon_i$ is the measurement error on the observed
(total) ellipticity of the $i^{th}$ galaxy. If the combined
astrophysical and measurement errors on the observed intrinsic position
angles are $\delta \alpha_i$, then
\be
\hat{\vn}_i = \left( \begin{array}{c} \sin(2\alpha^{\rm int}_i +
  2\delta\alpha_i) \\ -\cos(2\alpha^{\rm int}_i +
  2\delta\alpha_i) \end{array} \right) 
 = \vn^\perp_i \cos2\delta\alpha_i + \vn^\|_i \sin2\delta\alpha_i.
\label{eqn:A3}
\ee
These can be substituted into equations~(\ref{eqn:Amatrix_def}) and
(\ref{eqn:bvec_def}) to give
\ba
\mA &=& \sum_i w_i \left[ \vn^\perp_i {\vn^\perp_i}^T \cos^2
2\delta\alpha_i + \vn_i^\| {\vn_i^\|}^T
\sin^22\delta\alpha_i \right.
+ \left. \left(\vn_i^\| {\vn_i^\perp}^T +
\vn_i^\perp {\vn_i^\|}^T\right)
\sin2\delta\alpha_i\cos2\delta\alpha_i
\right], \label{eqn:A4} \\
\vb &=& \mA \vgamma + \delta \vepsilon + \delta \vb, \label{eqn:A5}
\ea
where
\ba
\delta \vepsilon &=& \sum_i w_i (\delta \vepsilon_i \cdot \hat{\vn}_i)
\hat{\vn}_i, \label{eqn:A6} \\
\delta \vb &=& \sum_i w_i \vepsilon_i \sin 2\delta\alpha_i
\left(\vn_i^\perp \cos 2\delta\alpha_i + \vn_i^\| \sin
2\delta\alpha_i\right). \label{eqn:A7}
\ea
Hence, from equations~(\ref{eqn:new_est}) and (\ref{eqn:A5}), the
estimator is given by
\be
\hat{\vgamma} = \vgamma + \mA^{-1} \delta \vepsilon + \mA^{-1} \delta
\vb.
\label{eqn:A8}
\ee
Ignoring the measurement errors on the galaxy ellipticities, $\delta
\vepsilon_i = 0$, and assuming small errors on the intrinsic position
angles, $\delta \alpha_i \ll 1$, to order $\delta\alpha^2$,
equation~(\ref{eqn:A8}) is
\be
\hat{\vgamma} = \vgamma + 2\sum_i w_i \, \delta\alpha_i^{\rm int}
\vepsilon_i \, \mA^{-1} \vn^\perp_i + \mathcal{O}(\delta\alpha^2),
\label{eqn:A9}
\ee
and the matrix, $\mA$ is
\be
\mA = \mA^\perp + \mathcal{O}(\delta\alpha) = \sum_i w_i \vn^\perp_i
{\vn^\perp_i}^T + \mathcal{O}(\delta\alpha).
\label{eqn:A10}
\ee
Therefore, ignoring $\mathcal{O}(\delta\alpha^2)$, we have
\be
\hat{\vgamma} - \vgamma \approx 2 \sum_i w_i \, \delta\alpha_i
\vepsilon_i \, (\mA^\perp)^{-1} \vn^\perp_i.
\label{eqn:A11}
\ee
Without loss of generality, we can assume uniform weights (normalized
such that $\sum_i w_i = 1$) and writing $\vm^\perp_i =
(\mA^\perp)^{-1} \vn^\perp_i$, equation~(\ref{eqn:A11}) is
\be
\hat{\vgamma} - \vgamma \approx \frac{2}{N} \sum_i \delta\alpha_i
\vepsilon_i \, \vm^\perp_i,
\label{eqn:A12}
\ee
where $N$ is the number of galaxies in the pixel. If we now assume
that $\delta\alpha_i$, $\vepsilon_i$ and $\vm^\perp_i$ are
independent, then we have 
\be
\hat{\vgamma} - \vgamma \approx 2 \left( \frac{1}{N}\sum_i \delta\alpha_i \right)
\left( \frac{1}{N}\sum_i \vepsilon_i \right) \left( \frac{1}{N}\sum_i
\vm^\perp_i \right).
\label{eqn:A13}
\ee
Taking the limit as $N \rightarrow \infty$, in the absence of an IA
signal, $\lgl \vm^\perp \rgl = \lgl \vepsilon \rgl = 0$, and
assuming the intrinsic position angle estimates are unbiased, $\lgl
\delta\alpha^{\rm int} \rgl = 0$. Thus, we have $\lgl \hat{\vgamma} \rgl =
\vgamma$ as required.  For $N$ finite, the standard error will be
given by
\be
\sigma_{\hat{\vgamma}} \approx 2 \frac{\lgl \delta\alpha^2 \rgl^{1/2} \lgl
\vepsilon^2 \rgl^{1/2} \lgl {\vm^\perp}^2 \rgl^{1/2}}{\sqrt{N}}.
\ee
For randomly distributed true position angles, we find that $\lgl
{\vm^\perp}^2 \rgl = 4$. Thus the error in the estimator is 
\be
\sigma_{\hat{\vgamma}} \approx 4 \frac{\alpha_{\rm rms} \, \vepsilon_{\rm rms}}{\sqrt{N}}. 
\ee

\label{lastpage}

\end{document}